\documentclass[aps,prx,twocolumn,superscriptaddress,showpacs]{revtex4}

\usepackage{amsmath}

\usepackage{amssymb}
\usepackage{graphicx,color}

\usepackage{amsmath,amsthm}
\usepackage{txfonts}

\usepackage{mathrsfs}

\usepackage{bm}

\newcommand{\Ref}[1]{(\ref{#1})}
\newcommand{\Z}{\mathbb{Z}}

\newcommand{\ccirc}{\kern0.2ex\vcenter{\hbox{$\scriptstyle\circ$}}\kern0.2ex}

\newcommand{\bA}{{\bar{A}}}

\def\be{\begin{eqnarray}}
\def\ee{\end{eqnarray}}

\newcommand{\cc}{\mathcal C}

\newcommand{\ch}{\mathcal H}

\newcommand{\cs}{\mathcal S}

\newcommand{\cv}{\mathcal V}

\newcommand{\fp}{\mathfrak{p}}

\newcommand{\g}{\gamma}

\newcommand{\eps}{\varepsilon}

\newcommand{\Sig}{\Sigma}

\renewcommand{\L }{\Lambda}

\renewcommand{\t}{\tau}

\newcommand{\rmd}{\mathrm d}

\newcommand{\lt}{\left}
\newcommand{\rt}{\right}

\newcommand{\rag}{\right\rangle}

\newcommand{\tr}{\mathrm{tr}}

\newcommand{\Ar}{\mathbf{Ar}}

\begin{document}

\sloppy

\title{\bf Discrete Gravity on Random Tensor Network and Holographic R\'enyi Entropy}

\author{Muxin Han}
\affiliation{Department of Physics, Florida Atlantic University, 777 Glades Road, Boca Raton, FL 33431, USA}
\affiliation{Institut f\"ur Quantengravitation, Universit\"at Erlangen-N\"urnberg, Staudtstr. 7/B2, 91058 Erlangen, Germany}

\author{Shilin Huang}
\affiliation{Institute for Interdisciplinary Information Sciences, Tsinghua University, Beijing 100084, China}

%\date{\small\today}

\begin{abstract}

In this paper we apply the discrete gravity and Regge calculus to tensor networks and Anti-de Sitter/conformal field theory (AdS/CFT) correspondence. We construct the boundary many-body quantum state $|\Psi\rangle$ using random tensor networks as the holographic mapping, applied to the Wheeler-deWitt wave function of bulk Euclidean discrete gravity in 3 dimensions. The entanglement R\'enyi entropy of $|\Psi\rangle$ is shown to holographically relate to the on-shell action of Einstein gravity on a branch cover bulk manifold. The resulting R\'enyi entropy $S_n$ of $|\Psi\rangle$ approximates with high precision the  R\'enyi entropy of ground state in 2-dimensional conformal field theory (CFT). In particular it reproduces the correct $n$ dependence. Our results develop the framework of realizing the AdS$_3$/CFT$_2$ correspondence on random tensor networks, and provide a new proposal to approximate the CFT ground state. 

\end{abstract}

\maketitle%\vspace{-7mm}

\section{Introduction}

\begin{figure}
\begin{center}
\includegraphics[width = 0.15\textwidth]{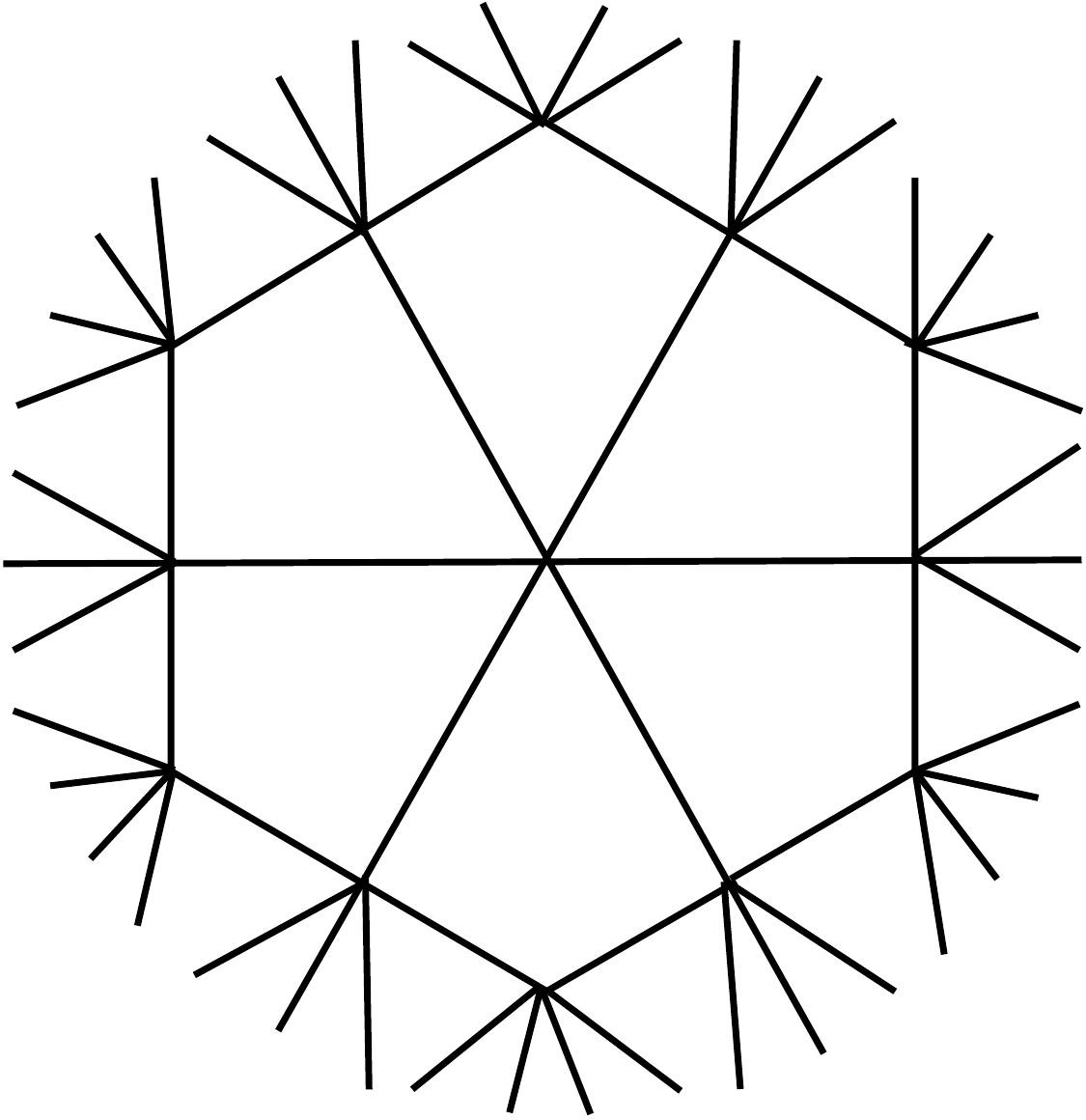}
\end{center}
\caption{An example of tensor network with rank-6 tensors. The tensor network state $|\Psi\rangle$ is given by an expansion with certain basis in the Hilbert space of many-body system $|\Psi\rangle=\sum_{\{a_i, \gamma_l\}}\prod_\fp (T_\fp)_{\gamma_1\, \gamma_2 \,...\, a_i...}|a_1,a_2,\cdots a_N\rangle$. The coefficients is constructed by distributing a rank-6 tensor $T_\fp$ at each 6-valent node $\fp$, such that each tensor index associates to a link adjacent to $\fp$. Connecting 2 nodes by a link means contracting the corresponding indices $\g_l$.}
\label{TNS}
\end{figure}

The tensor network is a quantum state of many-body system constructed by contracting tensors according to a network graph with nodes and links (FIG.\ref{TNS}). It is originated in condense matter physics because tensor network states efficiently compute ground states of many-body quantum systems \cite{Bridgeman:2016dhh,2015PhRvL.115r0405E}. In addition, the tensor network has wide applications to quantum information theory by its relation to error correcting codes and quantum entanglement \cite{TNandQEC}. It recently relates to quantum machine learning \cite{2016arXiv160505775M}, as well as neuroscience \cite{DBLP:journals/corr/NovikovPOV15}.

One of the fascinating developments of the tensor network is the recent relation to the AdS/CFT correspondence and emergent gravity program \cite{Orus:2014poa,Swingle:2009bg}. The AdS/CFT correspondence proposes that the quantum gravity theory on $d$-dimensional Anti-de Sitter (AdS) spacetime is equivalent to a conformal field theory (CFT) living at the $(d-1)$-dimensional boundary of AdS. It offers a dictionary between the observables of the $d$-dimensional bulk gravity theory and those of the $(d-1)$-dimensional boundary CFT. Properties of the bulk gravity and geometry may be reconstructed or emergent from the boundary CFT, known as the emergent gravity program \cite{VanRaamsdonk:2016exw,Verlinde:2016toy}. As an important ingredient of AdS/CFT, the bulk geometry relates holographically to the entanglement in the boundary CFT, via the Ryu-Takayanagi (RT) formula
 \be
S_{EE}({A}) = \frac{\Ar_{\textrm{min}}}{4G_N},\label{RT}
\ee
which identifies the entanglement entropy $S_{EE}({A})$ of a $(d-1)$-dimensional boundary region ${A}$ with the area $\Ar_{\textrm{min}}$ of the bulk $(d-2)$-dimensional minimal surface anchored to ${A}$ \cite{Ryu:2006bv,Lewkowycz:2013nqa,dche,Casini:2011kv,Hung:2011nu,Dong:2016fnf}. $G_N$ is the Newton constant in $d$ dimensions. $S_{EE}({A})$ satisfying RT formula is referred to as the holographic entanglement entropy (HEE) \cite{Rangamani:2016dms,deBoer:2013vca,Chen:2016kyz,Dong:2016wcf,Song:2016gtd,Ammon:2013hba,Ling:2015dma}

The tensor network can be understood as a discrete version of the AdS/CFT correspondence \cite{Hu:2017rsp,Pastawski:2015qua,Qi1,Miyaji:2016mxg,Bhattacharyya:2016hbx,Czech:2016nxc,2016arXiv161101140L,Li:2017qwu}. The tensor network states approximates the CFT state at the boundary, while the structure of tensor network emerges an bulk dimension built by layers of tensors. Tensors in the tensor network correspond to local degrees of freedom in the bulk \cite{Qi2,QiYY,Bhattacharyya:2017aly}. The architecture of the TN may be viewed as a process of real-space renormalization, such as multiscale entanglement renormalization ansatz (MERA), where the renormalization scale relates to the coordinate of the emergent dimension \cite{2015PhRvL.115r0405E,Swingle:2009bg,2008PhRvL.101k0501V,Vidal:2007hda}. The feature of tensor network makes it an interesting tool for realizing the AdS/CFT correspondence constructively from many-body quantum states. Among many recent progress, an important result is to reproduce the HEE on the tensor network.

The optimized scale-invariant MERA tensor networks are able to approximate ground states of (discreted) CFTs e.g. the critical Ising and Potts models. It is observed that the entanglement entropy of optimized MERA is proportional to the number of links on a minimal cut through the tensor network, as a nice realization of the RT formula \cite{Orus:2014poa,EV2011,Swingle:2009bg}. Moreover there has been two recent interesting exactly-solvable models that realize the RT formula on tensor networks, including \cite{Pastawski:2015qua} with tensor networks with perfect tensors, and \cite{Qi1} with random tensor networks. Given a boundary region $A$ containing a number of open links of the tensor network, the entanglement entropy of the tensor network in each of the above approaches reproduces an analog of the RT formula (See e.g. \cite{May:2016dgv,Li:2016eyr,Cotler:2017erl,Chirco:2017vhs,Pastawski:2016qrs,Peach:2017npp,Roberts:2016hpo,Hosur:2015ylk,Smolin:2016edy} for some more recent developments)
\be
S_{EE}({A})=\mathrm{Min}(\#_{cut})\cdot \ln D,\label{SEETN}
\ee 
where $\mathrm{Min}(\#_{cut})$ is the minimal number of tensor network links cut by a surface anchored to $A$. $D$ is the bond dimension (range of tensor index). Eq.\Ref{SEETN} is in general an upper bound in the tensor network, which is saturated in the case of perfect or random tensor network. The random tensor network approach has been related to the quantum geometry in loop quantum gravity (LQG) (\cite{review,review1,book,rovelli2014covariant} for reviews), which relates Eq.\Ref{SEETN} to the geometrical RT formula Eq.\Ref{RT} \cite{HanHung}.

Although the perfect and random tensor network approaches give exactly solvable models and have nice features, it is known that both approaches suffer the issue of flat entanglement spectrum. Although the entanglement (Von Neumann) entropy Eq.\Ref{SEETN} is consistent with the RT formula, R\'enyi entropies $S_n(A)$ from both approaches are all identical to Eq.\Ref{SEETN} with trivial $n$ dependence. But the (generalized) RT formula of R\'enyi entropy has a nontrivial $n$ dependence since the CFT R\'enyi entropy does \cite{Dong:2016fnf}. For instance, in any 2d CFT (CFT$_2$), the ground state has the universal R\'enyi entropy \cite{Calabrese:2004eu}
\be
S_n(A)=\lt(1+\frac{1}{n}\rt)\frac{c}{6} \ln \lt(\frac{l_A}{\delta}\rt),\label{CFTS}
\ee
where $c$ is the central charge, $l_A$ is the length of the region $A$, and $\delta$ is a UV cut-off. The CFT R\'enyi entropy manifestly has a nontrivial $n$ dependence. 

The mismatch of R\'enyi entropy implies that both tensor network states in \cite{Pastawski:2015qua,Qi1} fail to approximate the CFT ground state. An idea of the reason behind it may be extracted from the AdS/CFT: In the AdS/CFT, the CFT ground state (at strong-coupling) is dual to the bulk semiclassical AdS spacetime geometry. However the tensor networks designed in \cite{Pastawski:2015qua,Qi1} only consider the geometry of a spatial slice in AdS, without any input about time evolution. In the continuum AdS/CFT context, the correct entanglement spectrum are obtained by considering the spacetime geometry, and taking into account the bulk dynamics given by the Einstein equation \cite{Lewkowycz:2013nqa,Dong:2016fnf,dche,Hung:2011nu}, while \cite{Pastawski:2015qua,Qi1} don't have any dynamical input. In this work, we take the above idea as a hint to improve the tensor network approach, in which we resolve the issue. %The issue of entanglement spectrum may relate to the issue of dynamical input on the tensor network in the bulk or at the boundary.

Note that the optimized MERA for the CFT can derive the correct R\'enyi entropies as well as the non-flat entanglement spectrum \cite{EV2011}. However the goal of this work is to finding analytically a generic family of tensor networks that provide the correct CFT R\'enyi entropies without going through any numerical optimization at all. Our results develop important analytic methods for the tensor network AdS/CFT.

In this work, we resolve the above R\'enyi entropy issue by having the bulk dynamics on random tensor network states. The dynamics relates tensor networks to the discrete gravity model known as Regge calculus. We construct the state $|\Psi\rangle$ which is proposed as an approximation to the CFT ground state. As is anticipated by our proposal, $|\Psi\rangle$ indeed reproduces correctly the RT formula and CFT ground state R\'enyi entropy $S_n$ with correct $n$ dependence. Our results develop the framework of the AdS/CFT correspondence on random tensor networks, and provide a new proposal to approximate the large-N CFT ground state.

As another interesting aspect of our work, we show that the random tensor network effectively relates the boundary R\'enyi entropy to the partition functions of bulk gravity on the branched cover spacetime. This relation has been a key assumption in the existing AdS/CFT R\'enyi entropy computations \cite{Dong:2016fnf,dche,Lewkowycz:2013nqa}, while our work derives this relation in the context of AdS$_3$/CFT$_2$. We also derive the duality between the boundary R\'enyi entropy and bulk cosmic brane (in AdS$_3$/CFT$_2$), which has been proposed as a conjecture in \cite{Dong:2016fnf}.

\begin{figure}
\begin{center}
\includegraphics[width = 0.2\textwidth]{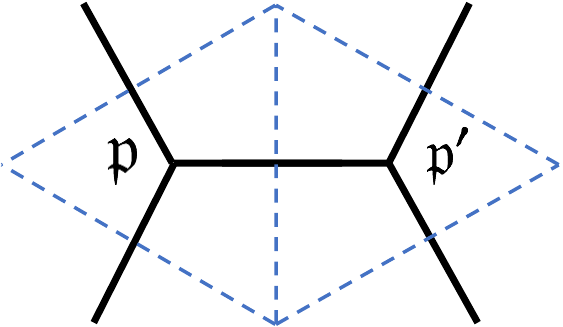}
\end{center}
\caption{A simple trivalent tensor network with 2 nodes and 5 links (4 open links). The tensor network is dual to a triangulation with 2 triangles.}
\label{NW}
\end{figure}

In this paper, we focus on 2d CFT and 3d bulk spacetime (AdS$_3$/CFT$_2$) in Euclidean signature. The CFT state $|\Psi\rangle$ is constructed by implementing bulk gravity dynamics to random tensor network states studied in \cite{QiYY}. The random tensor network constructed in \cite{QiYY} has random tensors at all nodes $\fp$, and has labels ${a}_{\fp,\fp'}$ on links $(\fp,\fp')$. Each ${a}_{\fp,\fp'}$ labels the non-maximal entangled state $|{a}_{\fp,\fp'}\rangle$ on each link. The tensor network is dual to a tiling of 2d spatial slice $\Sig$ (FIG.\ref{NW}). The entanglement entropy of $|{a}_{\fp,\fp'}\rangle$ relates to the length $L_\ell$ of the edge $\ell$ intersecting $(\fp,\fp')$ in the tiling. Thus each random tensor network as boundary CFT state, denoted by $|\vec{a}\rangle$, determines a set of edge lengths $L_\ell$ in the bulk. When the tiling is a triangulation, edge lengths uniquely determines a discrete geometry on $\Sig$ which approximates the smooth geometry as the triangulation is refined \cite{BARRETT1994107}. We make linear combinations of $|\vec{a}\rangle$ to write CFT states as 
\be
|\Psi\rangle=\sum_{\vec{a}}\Phi(\vec{a})|\vec{a}\rangle.\label{Psi0}
\ee
where the coefficients $\Phi(\vec{a})$ can be understood as a wave function of bulk geometry, by the relation between $\vec{a}$ and bulk edge lengths. Eq.\Ref{Psi0} is a holographic mapping from the bulk state $\Phi$ to the boundary state $\Psi$.

\begin{figure}
\begin{center}
\includegraphics[width = 0.4\textwidth]{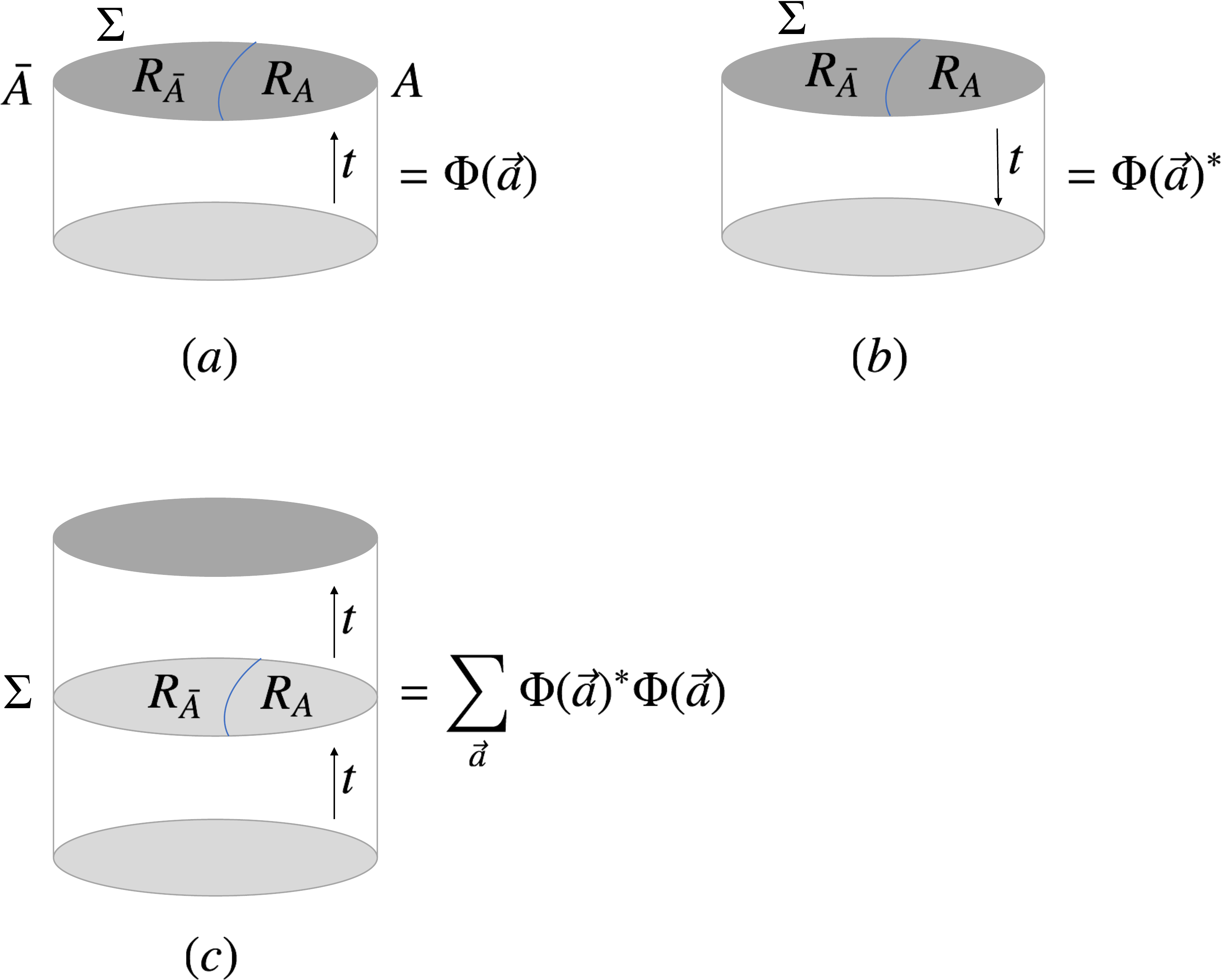}
\end{center}
\caption{(Triangulated) 3-manifolds $M$ in (a), $\bar{M}$ in (b), and $M_1$ in (c). The tensor network link variables $\vec{a}$ are the boundary data in the spatial slice $\Sig$, while the boundary data at other boundaries (including the infinitely past/future) doesn't affect our discussion. }
\label{trace}
\end{figure}

As mentioned above, the CFT ground state is expected dual to a physical state in the bulk corresponding to the semiclassical AdS spacetime geometry. But the link labels $\vec{a}$ in $\Phi(\vec{a})$ only relate to the geometry of 2d spatial slice $\Sig$. To let $\Phi(\vec{a})$ encode spacetime geometry, we propose $\Phi(\vec{a})$ to be the Wheeler-deWitt wave function. Namely, $\Phi(\vec{a})$ is a path integral of (Euclidean) Einstein gravity on spacetime $M$ whose boundary contains $\Sig$ (FIG.\ref{trace}(a)). The geometry on $\Sig$ determined by $\vec{a}$ is the boundary condition of the path integral. We construct $\Phi(\vec{a})$ to sum all possible bulk spacetime geometries satisfying the boundary condition, while in the semiclassical limit, it localizes at the classical AdS spacetime in 3d. The semiclassical limit relates to the large bond dimension of tensor network. 

Since $\vec{a}$ are data of discrete geometry, $\Phi(\vec{a})$ is the discrete version of Wheeler-deWitt wave function: It is a path integral of Regge calculus. Regge calculus is a discretization of Einstein gravity by triangulating spacetime geometries \cite{regge}. The discrete spacetime geometry is given by the edge lengths in the triangulation, known as the Regge geometry. Regge geometries converge to smooth geometries as triangulations are refined \cite{BARRETT1994107}. $\Phi(\vec{a})$ is a sum over all (Euclidean) Regge geometries on the spacetime $M$, weighted by the exponentiated Einstein-Regge action \cite{FFLR1}. The detailed explanations of $\Phi(\vec{a})$ and random tensor networks are presented in Section \ref{WDW}.

\begin{figure}
\begin{center}
\includegraphics[width = 0.45\textwidth]{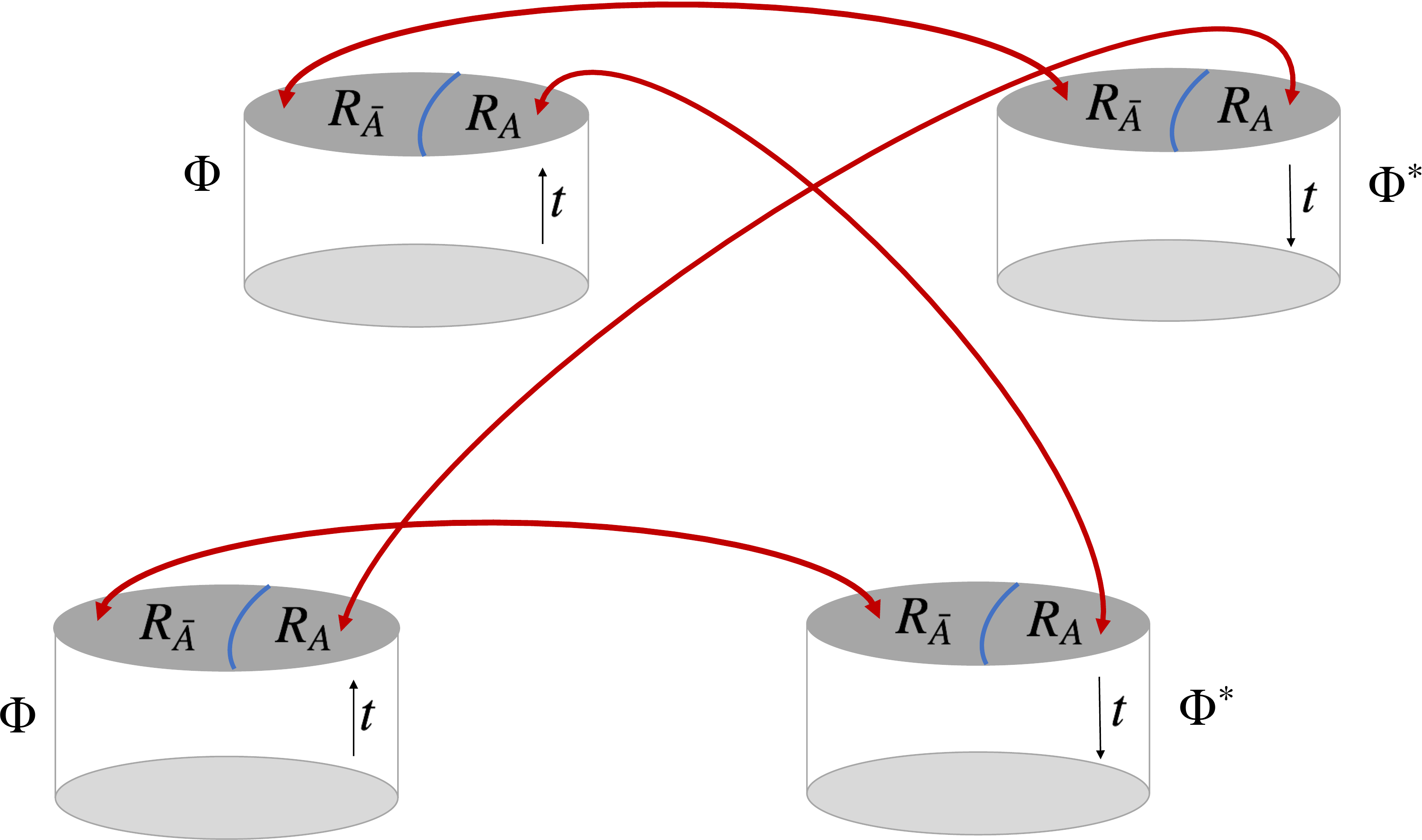}
\end{center}
\caption{The (triangulated) manifold $M_n$ ($n=2$) made by gluing $2n$ copies of $M$.% (b) The illustration of $M_2$ by suppressing 1 dimension. $R_A$ is a branch cut in $M_n$. 
} 
\label{replica}
\end{figure}

The R\'enyi entropy $\overline{S_n(A)}$ of $|\Psi\rangle$ at arbitrary $n\geq 2$ is computed in Section \ref{Renyi}. The computation involves averages of the random tensors at nodes $\fp$ in the tensor networks \cite{Qi1}. It turns out that the random average effectively glues $2n$ copies of path integrals $\Phi(\vec{a})$ ($\Phi(\vec{a})^*$) on $M$, and relate $\overline{S_n(A)}$ to the path integrals of gravity on branch cover spacetimes made by $2n$ copies of $M$. We derives that in the bulk semiclassical limit equivalent to large bond dimensions in the tensor networks,  
\be
\overline{S_n(A)}\simeq \frac{1}{1-n}\lt[I_{Bulk}(M_n)-n I_{Bulk}(M_1)\rt],\label{duality0}
\ee
where $I_{Bulk}(M_n)$ is the on-shell gravity action evaluated at the bulk solution on the branch cover manifold $M_n$ (FIG.\ref{replica}). The bulk solution has the $Z_n$ replica symmetry. Eq.\Ref{duality0} has been an assumption in AdS/CFT derivations of HEE in e.g. \cite{Lewkowycz:2013nqa,Dong:2016fnf,dche,Hung:2011nu}. But it is now derived from $|\Psi\rangle$ and random tensor networks. The derivation also gives the duality between the boundary R\'enyi entropy and bulk cosmic brane (in AdS$_3$/CFT$_2$), which has been proposed as a conjecture in \cite{Dong:2016fnf}. As a result, we show that $\overline{S_n(A)}$ reproduces the RT formula for holographic R\'enyi entropy for 2d CFT (Hung-Myers-Smolkin-Yale formula in \cite{Hung:2011nu})
\be
\overline{S_n(A)}\simeq \lt(1+\frac{1}{n}\rt)\frac{\Ar_{\textrm{min}}}{8G_N}.
\ee
Here $\Ar_{\textrm{min}}$ is the geodesic length in AdS$_3$. The above result of $\overline{S_n(A)}$ gives the R\'enyi entropy  Eq.\Ref{CFTS} of CFT$_2$ ground state with correct $n$ dependence. Our result is valid in the regime of large CFT central charge.

%Section \ref{fluctuation} analyzes the bound on the fluctuation of R\'enyi entropy from the random average value $\overline{S_n(A)}$, which shows in the bulk semiclassical limit the fluctuation is generically small. 

This work applies discrete geometry method such as Regge calculus to study tensor networks (see \cite{Gubser:2016htz} for other application of discrete gravity in AdS/CFT). $|\Psi\rangle$ encodes the dynamics of bulk geometries, which is given by the discrete Einstein equation. It is interesting to further understand how the bulk dynamics might relate to the dynamics of boundary CFT, and whether a boundary CFT Hamiltonian might be induced from the bulk dynamics. It is interesting to compare our proposal of CFT ground state $|\Psi\rangle$ to the MERA approach, to extend our result to finite central charges and understand the relation between $|\Psi\rangle$ and concrete many-body systems such as critical spin-chains. The understanding of these aspects should develop tensor network models to realize the AdS/CFT correspondence at the dynamical level. The research on these aspects is currently undergoing.

% the purpose is to design a state mimic the ground state of CFT in the perspective of R\'enyi entropies with the right n dependence. 

\section{Random Tensor Network and Wheeler-deWitt Wave Function}\label{WDW}

In this work we consider trivalent random tensor network states. A tensor network is viewed as a discrete 2d spatial slice $\Sig$ of 3d bulk spacetime. It is made by a large number of trivalent random tensors $|\cv_\fp\rangle\in \ch^{\otimes 3}\equiv\ch_\fp$ at each tensor network node $\fp$. The Hilbert space $\ch$ is of dimension $D$. We decompose $\ch$ into a number of subspaces $\ch\simeq\oplus_a V_a$ and denote $\dim(V_a)\equiv d[a]$. Each internal link $(\fp,\fp')$ of the tensor network associates with a maximal entangled state in $V_a\otimes V_a$ of certain $a$, 
\be
|a_{\fp,\fp'}\rangle%\equiv\sum_{\mu,\nu}a_{\mu\nu}|\mu\rangle_\fp\otimes|\nu\rangle_{\fp'}
=\frac{1}{\sqrt{d[a]}}\sum_{\mu=1}^{d[a]}|\mu\rangle_\fp\otimes|\mu\rangle_{\fp'}\label{a}
\ee
where $|\mu\rangle_\fp$ is a basis in $\ch$. It satisfies $\langle a_{\fp,\fp'}|b_{\fp,\fp'}\rangle=\delta_{ab}$. A class of random tensor networks $|\vec{a}\rangle$ can be defined by the (partial) inner product between $|\cv_\fp\rangle$ at all $\fp$ and $|a_{\fp,\fp'}\rangle$ on all internal links
\be
|\vec{a}\rangle=\otimes_{\fp,\fp'}\langle a_{\fp\fp'}|\otimes_\fp|\cv_\fp\rangle.\label{PhiEV}
\ee
The inner product takes place in $\ch$ at each end point $\fp$ or $\fp'$ of each link. $|\vec{a}\rangle$ is a state in the boundary Hilbert space $\ch^{N_\partial}$, where $N_\partial$ is the number of open links.

The label $\vec{a}$ relates to the amount of entanglement on each internal link $(\fp,\fp')$. The entanglement entropy $S(|a_{\fp,\fp'}\rangle)$ of $|a_{\fp,\fp'}\rangle$ is $\ln d[a_{\fp,\fp'}]$, where $d[a_{\fp,\fp'}]$ is effectively the bond dimension on $(\fp,\fp')$ in $|\vec{a}\rangle$.

The above class of tensor network states is proposed in \cite{QiYY}. We consider the boundary state $|\Psi\rangle$ as a linear combination \footnote{The set of $|\vec{a}\rangle$ are shown to form an overcomplete basis in the boundary Hilbert space if there is no restriction on the link variables $\vec{a}$ \cite{QiYY}. Whether $|\vec{a}\rangle$ with $\vec{a}$ as tiling of the spatial slice still form an over-complete basis is an interesting question but doesn't affect the present analysis.}
\be
|\Psi\rangle=\sum_{\vec{a}}\Phi(\vec{a})|\vec{a}\rangle.\label{Psi}
\ee

Here we understand the trivalent tensor network to be dual to a triangulation of the spatial slice $\Sig$ (FIG.\ref{NW}). Namely, each node $\fp$ located at the center of a triangle $\Delta_\fp$ in the triangulation. Each link $(\fp,\fp')$ intersects transversely an internal edge $\ell$ shared by 2 triangles $\Delta_\fp,\Delta_{\fp'}$. Open links in tensor network intersect transversely the edges at the boundary of triangulation.

The label $\vec{a}$ is understood as the discrete geometry in the bulk of $\Sig$ \cite{QiYY}, in the sense that the edge length $L_\ell$ of $\ell$ intersecting $(\fp,\fp')$ is proportional to the entanglement entropy $S(|a_{\fp,\fp'}\rangle)$ on each link:
\be
L_\ell\equiv 4\ell_P\ln d[a_{\fp,\fp'}]\label{length}
\ee
Here $\ell_P=G_N\hbar$ is the Planck length in 3d. In this proposal, the bulk geometry is understood as emergent from the entanglement in tensor network state. The relation can be obtained from the recent proposal of understanding tensor networks as the effective theory from coarse graining quantum gravity at Planck scale \cite{HanHung}, in which one derives that the bond dimension $d[a_{\fp,\fp'}]$ of tensor network $|\vec{a}\rangle$ satisfies $d[a_{\fp,\fp'}]\simeq e^{L_\ell/4\ell_P}$.

By the relation between $\vec{a}$ and bulk geometry, Eq.\Ref{Psi} is a boundary state by summing over all bulk spatial geometry on $\Sig$, while $\Phi(\vec{a})$ is a wave function of bulk geometry. Eq.\Ref{Psi} defines a holographic mapping from the bulk states of geometry to the boundary states of CFT. We propose the following boundary state $|\Psi\rangle$ whose bulk wave function $\Phi(\vec{a})$ (pre-image of the holographic mapping) is an Wheeler-deWitt wave function in 3d Euclidean gravity. Namely $\Phi(\vec{a})$ is a path integral of gravity on a 3d solid cylinder $M$, whose boundary includes $\Sig$ in addition to the boundary where CFT lives (FIG.\ref{trace}(a)). The geometry $\vec{a}$ is the boundary condition on $\Sig$ in the path integral. The path integral may also depend on the boundary conditions at other boundaries of $M$. But we make those boundary conditions implicit since they play no role in the following analysis.

Since $\vec{a}$ gives a discrete geometry with a set of edge lengths $L_\ell$, more precisely, $\Phi(\vec{a})$ is a discrete version of the Wheeler-deWitt wave function. Indeed, we consider a sufficiently refined triangulation of $M$, and impose discrete metrics on the triangulation. Namely each tetrahedron in the triangulation carries a 3d hyperbolic geometry with constant curvature $-{L_{AdS}^{-2}}$. Tetrahedron edges are geodesics in the hyperbolic space, and have edge lengths $L_\ell$. The set of edge lengths $\{L_\ell\}$ on the triangulation defines a discrete metric of Regge geometry \cite{regge,FFLR1,BD1}. We define $\Phi(\vec{a})$ to be a path integral of discrete gravity on the triangulation by summing over all $L_\ell$ in the bulk of $M$
\be
\Phi(\vec{a}):=\sum_{L_\ell} e^{-S_{Regge}(M)} \label{Phi}
\ee
The boundary condition at $\Sig$ is $L_{\ell\subset\Sig}=4\ell_P\ln d[a_{\fp,\fp'}]$. We use $\sum_{L_\ell}$ instead of integration because $L_\ell$ are assumed as discrete data, to be consistent with $L_{\ell\subset\Sig}$. $S_{Regge}(M)$ is the Regge action of Euclidean gravity on the triangulated 3-manifold $M$ evaluated at the discrete metric $\{L_\ell\}$:
\be
S_{Regge}(M)=-\frac{1}{8\pi\ell_P}\lt[\sum_{\ell\subset\text{bulk}(M)}L_\ell\,\eps_\ell+\sum_{\ell\subset\partial M}L_\ell\,\Theta_\ell-\frac{V(M)}{L_{AdS}^2}\rt].\nonumber
\ee
$L_{AdS}$ relates to $\ell_P$ and the central charge of CFT by ${L_{AdS}}=\frac{2}{3}c\ell_P$%\cite{Henningson:1998gx}
. In the following, the large bond dimension limit $d[a_{\fp,\fp'}]\gg 1$ will be taken in the computation of Renyi entropies. The limit corresponds to the bulk semiclassical limit $\ell_P\to0$ and large central charge $c\gg1$ as both $L_{\ell}$ and $L_{AdS}$ are kept fixed. $\eps_\ell$ is the bulk deficit angle hinged at the bulk edge $\ell$. $\eps_\ell$ is a discretization of the bulk curvature. Each bulk edge $\ell$ is shared by a number of tetrahedra $t$. In each $t$, the dihedral angle between 2 faces joint at $\ell$ is denoted by $\theta(t,\ell)$. The deficit angle is defined by
\be
\eps_\ell=2\pi-\sum_{t,\,\ell\subset t}\theta(t,\ell),\quad \ell\subset\text{bulk}.
\ee
Each boundary edge $\ell$ is shared by 2 boundary triangles. $\Theta_\ell$ is the angle between their outward pointing normals, equivalently  
\be
\Theta_\ell=\pi-\sum_{t,\,\ell\subset t}\theta(t,\ell), \quad \ell\subset\text{boundary}.
\ee
$\Theta_\ell$ relates to the boundary extrinsic curvature. The 1st term in $S_{Regge}$ is the discretization of Ricci scalar term of Einstein-Hilbert action, while the 2nd term is the discretization of Gibbons-Hawking boundary term \cite{Hartle1981}. The last term is the cosmological constant term where $V(M)$ is the total volume of $M$. All quantities $\eps_\ell,\ \Theta_\ell$, and $V(M)$ are determined by edge lengths $L_\ell$.  

In order to be the boundary condition of Regge geometry, $L_{\ell\subset\Sig}=4\ell_P\ln d[a_{\fp,\fp'}]$ have to be the edge lengths of hyperbolic triangles, which triangulate $\Sig$. $L_{\ell\subset\Sig}$ have to be a discrete metric of $\Sig$, which constrains the allowed data $\vec{a}$ entering the sum in Eq.\Ref{Psi}.

%Note that the definition of $\Phi(\vec{a})$ involves the length scale $\ell_P$ in order to make Regge action dimensionless. 

Applying $\Phi(\vec{a})$ in Eq.\Ref{Phi} to the holographic mapping Eq.\Ref{Psi}, we obtain a boundary CFT state $|\Psi\rangle$, and we propose the resulting $|\Psi\rangle$ to be the ground state of the boundary large-N CFT, in the bulk semiclassical regime $\ell_P\ll L_\ell$ equivalent to the large bond dimension limit by Eq.\Ref{length}. As $\ell_P\ll L_\ell$, the path integral $\Phi(\vec{a})$ localizes at the solution of equation of motion (deriving equation of motion uses the Schl\" afli identity of hyperbolic tetrahedra $-{\delta V(t)}/{L_{AdS}^2}=\sum_{\ell\subset t}L_\ell\delta \theta(t,\ell)$, see e.g. \cite{BD1})
\be
\eps_\ell=0,\quad \forall\ \ell\subset \mathrm{bulk}({M}).\label{EOMregge}
\ee
which is the discretized Einstein equation in 3d. Vanishing $\eps_\ell$ everywhere means that the 3d Regge geometry is a smooth Euclidean AdS$_3$. So $\Phi(\vec{a})$ is a semiclassical wave function of bulk AdS$_3$ geometry. The holographic mapping is expected to map the bulk semiclassical state of AdS$_3$ to the ground state of boundary CFT$_2$. 

In Section \ref{Renyi}, we check our proposal by computing the R\'enyi entropies $S_n$ of the state $|\Psi\rangle$. We show that $|\Psi\rangle$ indeed reproduces correctly the R\'enyi entropies of CFT ground state with the correct $n$ dependence, in the regime $\ell_P\ll L_\ell$.

Before we come to the discussion of the R\'enyi entropy. There is a remark about the Regge calculus that we employ above: The Regge calculus with cosmological constant $\L$ has 2 formulations, depending on the curvature of tetrahedra in Regge geometries \cite{BD,BD1,HHKR}. One may use flat tetrahedra in the same way as the Regge calculus without $\L$, in order to have simpler Regge geometries. But the disadvantage is that the constant curvature space is an approximation rather than exact solution. From this perspective, it is more convenient for us to employ the second formulation with constant curvature tetrahedra, where the constant curvature is consistent with the cosmological constant $\L$. Consequently, the Regge equation of motion Eq.\Ref{EOMregge} gives the smooth AdS$_3$ geometry as the exact solution, although we start with a discrete gravity theory.

\section{R\'enyi Entropies}\label{Renyi}

We compute R\'enyi entropies $S_n$ of the state $|\Psi\rangle$ by specifying a boundary region $A\subset\partial\Sig$ which contains a subset of open links. Recall that $|\Psi\rangle$ is made by random tensors at nodes, the $n$-th R\'enyi entropy is given by an average over random tensors \cite{Qi1} %(the computation uses the technique developed in \cite{Qi1})
\be
\overline{S_n(A)}=\frac{1}{1-n}\ln\frac{\overline{\tr(\rho_A^n)}}{\overline{(\tr\rho_A)^n}}.\label{Sn}
\ee
The fluctuation away from the average is discussed in Appendix \ref{fluctuation}. $\rho_A$ is the reduced density matrix by tracing out the degrees of freedom located in the complement $\bar{A}=\partial\Sig\setminus A$. $\tr(\rho_A^n)$ can be conveniently written in terms of the pure density matrix $\rho=|\Psi\rangle\langle\Psi|$ 
\be
\tr(\rho_A^n)%=\langle \mu_\ell^{(1)}|\rho_A|\mu^{(2)}_\ell\rangle_A\ \langle \mu^{(2)}_\ell|\rho_A|\mu_\ell^{(3)}\rangle_A\cdots\langle \mu^{(n)}_\ell|\rho_A|\mu_\ell^{(1)}\rangle_A
=\tr\lt[(\rho\otimes\cdots\otimes\rho)\cc^{(n)}_A\rt],\label{tracehon1}
\ee
where the trace is taken in $n$ copies of boundary Hilbert space. $\cc^{(n)}_A$ cyclicly permutes the states of region $A$, leaving the states of $\bA$ invariant, e.g. for $n=2$, $\cc^{(2)}_A(|\mu_\ell^{(1)}\rangle_A|\mu^{(1)}\rangle_\bA\otimes|\mu^{(2)}\rangle_A|\mu^{(2)}\rangle_\bA)
=|\mu_\ell^{(2)}\rangle_A|\mu^{(1)}\rangle_\bA\otimes|\mu^{(1)}\rangle_A|\mu^{(2)}\rangle_\bA$ where $|\mu\rangle$ forms a basis in the boundary Hilbert space. 

Define the pure state density matrix $\rho_P=| E_{\vec{a},\Phi}\rangle\langle E_{\vec{a},\Phi}|$ where $| E_{\vec{a},\Phi}\rangle=\sum_{\vec{a}}\Phi(\vec{a})\otimes_{\fp,\fp'}| a_{\fp\fp'}\rangle$. $\tr(\rho_A^n)$ can be written as
\be
\tr(\rho_A^n)=\tr\lt[\lt(\rho_P^{\otimes n}\otimes_\fp|\cv_\fp\rangle\langle \cv_\fp|\rt)^{\otimes n}\cc^{(n)}_A\rt].\label{tracerhon}
\ee
where the trace is taken in all $\ch_\fp\equiv \ch^{\otimes 3}$ at all nodes. $\tr(\rho_A^n)$ contains $2n$ copies of wave functions $\Phi(\vec{a})$ or $\Phi(\vec{a})^*$.

The random average $\overline{\tr\rho_A^n}$ relates to average $n$ copies of random tensors $|\cv_\fp\rangle\langle\cv_\fp|$ at each node $\fp$. Taking an arbitrary reference state $|0_\fp\rangle\in\ch_\fp$, the random tensor $|\cv_\fp\rangle=U_\fp|0_\fp\rangle$ with $U_\fp$ unitary transformation on $\ch^{\otimes 3}\equiv \ch_\fp$. The Haar random average is given by \cite{Qi1,church}:
\be
\overline{\lt(|\cv_\fp\rangle\langle \cv_\fp|\rt)^{\otimes n}}&=&\int\rmd U_\fp \lt(U_\fp|0_\fp\rangle\langle 0_\fp|U_\fp^{\dagger}\rt)^{\otimes n}\nonumber\\
&=&\frac{1}{C_{n,\fp}}\sum_{g_\fp\in\text{Sym}_n}g_{\fp}\in\ch_\fp^{\otimes n}\otimes\ch_\fp^*{}^{\otimes n}\label{random}
\ee 
where $\rmd U_\fp$ is the Haar measure on the group of all unitary transformations. $\sum_{g_\fp\in\text{Sym}_n}$ sums over all permutations $g_\fp$ acting on $\ch_\fp^{\otimes n}$. The overall constant $C_{n,\fp}=\sum_{g_\fp\in\text{Sym}_n}\tr g_{\fp}=(\dim\ch_\fp+n-1)!/(\dim\ch_\fp-1)!$.

Inserting this result in $\tr\rho_A^n$, the average $\overline{\tr\rho_A^n}$ becomes a sum over all permutations $\{g_\fp\}$ at all nodes $\fp$, where each term associates to a choice of $g_\fp$ at each $\fp$. It is straight-forward to compute that for large bond dimension $D\gg1$, the sum over $\{g_\fp\}$ is dominated by the contribution from $\{g_\fp\}$ satisfying the following boundary condition: $g_\fp=I$ if $\fp$ is adjacent to $\bA$, while $g_\fp=(\cc^{(n)})^{-1}$ if $\fp$ is adjacent to $A$ \cite{Qi1}.

Given $\{g_\fp\}$ satisfying the boundary condition, $\{g_\fp\}$ contains different domains on $\Sig$ with different permutations. We denote by $R_{g}$ the closed region in which $\fp\in R_{g}$ are of constant $g_\fp=g$. $R_g\cap R_{g'}\equiv\cs_{g,g'}$ denotes the domain wall shared by $R_g, R_{g'}$ with two different permutations $g\neq g'$.

Recall that the Wheeler-deWitt wave functions $\Phi(\vec{a})$ is a path integral on a triangulated manifold $M$, where $\vec{a}$ is the boundary condition. The product of $2n$ copies of $\Phi(\vec{a})$ or $\Phi(\vec{a})^*$ in $\tr(\rho_A^n)$ is the path integral on the product of $2n$ copies of manifolds $M$ or $\bar{M}$ (FIG.\ref{trace}) with identical triangulations. The result of random average Eq.\Ref{random} permutes the boundary conditions $\vec{a}$ and identifies boundary conditions $\vec{a}$. The trace in Eq.\Ref{tracerhon} effectively glue the $2n$ path integrals. As a result, the sum over $\{g_\fp\}$ becomes a sum of path integrals on different manifolds. Each $\{g_\fp\}$ determines a gluing of $2n$ copies of $M$ and $\bar{M}$ in certain manner. The path integral is defined on the resulting manifold. See FIG.\ref{replica} for an example of $n=2$. A graphic computation of $\tr(\rho_A^n)$ is presented in Appendix \ref{compute}.

Here is a description of the manifold determined by each $\{g_\fp\}$: Inside a domain $R_g$, the $i$-th copy of $\bar{M}$ is glued to the $g(i)$-th copy of $M$. In a neighboring domain $R_{g'}$, the $i$-th copy of $\bar{M}$ is glued to the $g'(i)$-th copy of $M$. Therefore the gluing in $R_g$ and $R_{g'}$ results in a branch cut, where the domain wall $\cs_{g,g'}$ is a 1d branch curve containing all branch points. Taking all domains with different permutations into account, each $\{g_\fp\}$ determines a manifold $M_{\{g_\fp\}}$ made by gluing $n$ copies of $M$ and $n$ copies of $\bar{M}$, with a number of branch cuts. FIG.\ref{replica} illustrates a simple situation with $n=2$, where there are 2 domains of the identity $I$ and cyclic $\cc^{(2)}$. The domain wall is a branch curve with $\Z_2$ symmetry. This situation generalizes to a general domain wall $\cs_{g,g'}$ and a cycle $c$ in $g^{-1}g'$. Indeed, each domain wall $\cs_{g,g'}$ becomes $\chi(g^{-1}g')$ branching curves $\cs_{g,g'}(c)$, where $\chi(g^{-1}g')$ is the number of cycles in $g^{-1}g'$. Each branch curve $\cs_{g,g'}(c)$ associates to a cycle $c\in C(g^{-1}g')$, and has a local $\Z_{n_c}$ symmetry, where $n_c$ is the length of the cycle $c$ (the number of involved elements $i\in c$), satisfying $\sum_c n_c=n$.  

At each branch curve $\cs_{g,g'}(c)$ where $2 n_c$ copies of $M$ and $\bar{M}$ meet, Regge actions in the $2 n_c$ copies of $\Phi({\vec{a}}),\Phi({\vec{a}})^*$ contribute boundary terms $L_\ell\Theta_\ell$ to each $\ell\subset\cs_{g,g'}(c)$. In addition, thanks to a normalization factor which contributes $(1-n_c)\ln d[a_{\fp,\fp'}(c)]=(1-n_c)\frac{L_\ell}{4\ell_P}$ (see Appendix \ref{compute}), the total contributions at each $\cs_{g,g'}(c)$ precisely make a new bulk term $\frac{1}{8\pi \ell_P}\sum_{\ell\subset\cs} L_\ell\eps_\ell$ of Regge action. 
%\be
%&&\frac{1}{4 \ell_P}\lt(1-n_c\rt)\sum_{\ell\subset\cs} L_\ell+\frac{1}{8\pi \ell_P}\sum_{\ell\subset\cs} L_\ell\lt(2\pi n_c-\sum_{t,\,\ell\subset t}\theta(t,\ell)\rt)\nonumber\\
%&=&\frac{1}{8\pi \ell_P}\sum_{\ell\subset\cs} L_\ell\lt(2\pi-\sum_{t,\,\ell\subset t}\theta(t,\ell)\rt)
%&=&\frac{1}{8\pi \ell_P}\sum_{\ell\subset\cs} L_\ell\eps_\ell
%\ee
On the other hand, it is easy to see that when $M$ glues to $\bar{M}$ inside a domain $R_g$, a pair of boundary terms from $S_{Regge}(M)$ and $S_{Regge}(\bar{M})$ again makes a bulk term of Regge action on the glued manifold \cite{Hartle1981}.

As a result, $\overline{\tr(\rho_A^n)}$ is written as a sum of discrete path integrals of Regge actions on different manifolds $M_{\{g_\fp\}}$:
\be
\overline{\tr(\rho_A^n)}\sim%\prod_\fp\frac{1}{C_{n,\fp}}\lt(D^n\rt)^{N_\partial}\sum_{\{g_\fp\}}
\sum_{\{L_\ell\}}e^{-S_{Regge}\lt(M_{\{g_\fp\}}\rt)}.\label{sum2}
\ee
It allows us to translate the R\'enyi entropy of boundary CFT to the bulk geometry.

As is mentioned above, we consider the regime $\ell_P\ll L_\ell$. On each $M_{\{g_\fp\}}$, the dominant contribution comes from the solution of equation of motion 
\be
\eps_\ell=0,\quad \forall\ \ell\subset \mathrm{bulk}({M}_{\{g_\fp\}})
\ee
It implies that as the leading contribution, the geometry on $M_{\{g_\fp\}}$ is smooth AdS$_3$ everywhere. The on-shell action gives
\be
\sum_{\{L_\ell\}}e^{-S_{Regge}\lt(M_{\{g_\fp\}}\rt)}\sim e^{-\frac{1}{8\pi \ell_P}\frac{V\lt(M_{\{g_\fp\}}\rt)}{L_{AdS}^2}+\text{boundary terms}}
\ee
%It turns out that the boundary term will cancel the similar contribution at the denominator in Eq.\Ref{Sn}. So in the following  discussion we often omit the boundary term. 
%where $\sum_{\{L_\ell\}_{\ell\subset\cs}}$ sums the freedoms the edge lengths within domain walls, i.e. it sums embeddings of branch curves in the hyperbolic geometry. 

Focus on a given $M_{\{g_\fp\}}$, locally at each $\cs_{g,g'}(c)$ of a given cycle $c$, the geometry has a local $\Z_{n_c}$ symmetry at both continuum level and discrete level, because we use the same triangulation on all copies of $M$ and $\bar{M}$. We cut a local neighborhood $N_{n_c}$ at $\cs_{g,g'}(c)$ from $M_{\{g_\fp\}}$, and make a $\Z_{n_c}$ quotient. The orbifold is denoted by $\hat{N}_{n_c}=N_{n_c}/\Z_{n_c}$. The geometry on $\hat{N}_{n_c}$ has a conical singularity at $\cs_{g,g'}(c)$ with deficit angle $2\pi\lt(1-\frac{1}{n_c}\rt)$. In the language of \cite{Dong:2016fnf}, the geometry we derive is back-reacted by a cosmic brane with tension $T_{n_c}=\frac{n_c-1}{4n_c \ell_P}$, located at $\cs_{g,g'}(c)$. We may analytic continue $n_c$ by considering arbitrary conical singularity or brane tension. 

The geometry of branch curve $\cs_{g,g'}(c)$ is determined by the equation of motion as in \cite{dche}. Both on-shell geometries on ${N}_{n_c}$ and $\hat{N}_{n_c}$ are AdS$_3$, except the conical singularity of $\hat{N}_{n_c}$. $\hat{N}_{n_c}$ is a fundamental domain in $N_{n_c}$ of $\Z_n$. $\hat{N}_{n_c}$ may be obtained by cutting $N_{n_c}$ into $n_c$ identical pieces, pick up one piece, followed by identifying its 2 cut boundaries. $\hat{N}_{n_c}$ is AdS$_3$ away from the singularity. So the glued boundaries can be chosen to be identical hyperbolic surfaces intersecting at the singularity. The singularity $\cs_{g,g'}(c)$ as the intersection has to be a geodesic (hyperbola) in the hyperbolic plane. The length $L_{\cs_{g,g'}(c)}(n_c)$ of $\cs_{g,g'}(c)$ explicitly depends on $n_c$. Since the triangulation of $M$ has been fixed, we only consider $\cs_{g,g'}(c)$ made by the edges in the triangulation. Otherwise the equation of motion cannot be satisfied and the domain wall $\cs_{g,g'}(c)$ doesn't give leading order contribution.  

Consider the volume of $\hat{N}_{n_c}$. We analytic continue $n_c$ and compute the derivative. By Schl\"afli identity of hyperbolic tetrahedra and keeping $\eps_\ell=0$ fixed in the bulk $\partial_{n_c} V\lt(\hat{N}_{n_c}\rt)
%=\sum_{\ell\subset \cs_{g,g'}(c)}L_\ell \partial_{n_c}\lt(\frac{2\pi}{n_c}\rt)
=\frac{2\pi L_{AdS}^2}{n_c^2}L_{\cs_{g,g'}(c)}(n_c)$. Integrating this relation gives
\be
{V\lt(\hat{N}_{n_c}\rt)} ={V\lt({N}_{1}\rt)}+\int_1^{n_c}\frac{2\pi {L_{AdS}^2}}{q^2}L_{\cs_{g,g'}(c)}(q)\,\rmd q,\label{volume0}
%=\L V\lt({N}_{1}\rt)+2\pi\sum_{\ell\subset \cs_{g,g'}(c)}L_\ell\lt[\frac{1}{n_c}-1\rt]
\ee
where $N_1$ has no singularity at $\cs_{g,g'}(c)$. 

Because the geometry is smooth AdS$_3$ on $\hat{N}_{n_c}$ away from $\cs_{g,g'}(c)$, to compute $L_{\cs_{g,g'}(c)}(q)$, we use the metric on $\hat{N}_{q}$ in the hyperbolic foliation \cite{Casini:2011kv,Hung:2011nu}:
\be
\rmd s^2=\lt(\frac{r^2}{L_{AdS}^2}-\frac{1}{q^2}\rt)L_{AdS}^2\rmd \t^2+\frac{\rmd r^2}{\frac{r^2}{L_{AdS}^2}-\frac{1}{q^2}}+r^2{\rmd u^2}.
\ee
The periodicity of $\t$ is $\t\sim\t+2\pi$. $r$ satisfies $r\geq L_{AdS}/q$. $\int \rmd u$ is the geodesics length in the hyperbolic plane with unit curvature. $\cs_{g,g'}(c)$ is located at the origin $r= L_{AdS}/q$. Rewriting the metric by $(2qL_{AdS})^{-1}\xi^2=r-L_{AdS}/q$ and considering the limit $\xi\to 0$ shows 
%\be
%\rmd s^2\sim \frac{\xi^2}{q^2}\rmd\t^2+\rmd\xi^2+\lt(\frac{\xi^2}{2qL_{AdS}}+\frac{L_{AdS}}{q}\rt)^2\rmd u^2,
%\ee
%which manifests the conical singularity at $\xi\to0$. 
the length of $\cs_{g,g'}(c)$ is given by $L_{\cs_{g,g'}(c)}(q)={q}^{-1}{L_{AdS}}\int_{\cs_{g,g'}(c)} \rmd u\equiv {q}^{-1}{L_{AdS}}l_{\cs_{g,g'}}$, where $l_{\cs_{g,g'}}$ is the geodesic length of $\cs_{g,g'}$ evaluated in the hyperbolic plane with unit curvature. $l_{\cs_{g,g'}}$ is independent of $c$. As a result,
\be
-\int_1^{n_c}\frac{2\pi}{q^2}L_{\cs_{g,g'}(c)}(q)\,\rmd q=\frac{1-n_c^2}{n_c^2}\pi{L_{AdS}}l_{\cs_{g,g'}}\label{26}
\ee

The volume of $N_c$: $V(N_{n_c})= n_c V\lt(\hat{N}_{n_c}\rt)$, i.e. $n_c$ times Eq.\Ref{volume0}. When we glue back $N_{n_c}$ in $M_{\{g_\fp\}}$. The first term in Eq.\Ref{volume0} gives $n_c V\lt({N}_{1}\rt)$, and effectively replaces $N_{n_c}$ by $n_c$ copies of ${N}_{1}$, which resolves the branch curve $\cs_{g,g'}(c)$ in $M_{\{g_\fp\}}$. When all branch curves are resolved, $M_{\{g_\fp\}}$ reduces to $n$ copies $M_1$. Therefore when we sum all domain walls and all cycles,  
\be
{V\lt(M_{\{g_\fp\}}\rt)}
= {n V\lt({M}_{1}\rt)}-{\pi  L_{AdS}^3}\sum_{\cs_{g,g'}}\sum_{c\in C(g^{-1}g')}\frac{1-n_c^2}{n_c}l_{\cs_{g,g'}}\nonumber
\ee
It is shown in Appendix \ref{DW} that the sum is dominant by $\{g_\fp\}$ with only a single domain wall $\cs$ separating $I$ in $R_\bA$ and $(\cc^{(n)})^{-1}$ in $R_A$, where $R_A$ (or $R_{\bA}$) is the region bounded by the boundary region $A$ (or $\bA$) and the domain wall (FIG.\ref{trace}). We denote the corresponding $M_{\{g_\fp\}}$ by $M_{n}$
\be
{V\lt(M_{n}\rt)}
= {n}{V\lt({M}_{1}\rt)}-{\pi  L_{AdS}^3}\frac{1-n^2}{n}l_{\cs}
\label{lnregge1}
\ee
The contribution of any other $M_{\{g_\fp\}}$ is much less than Eq.\Ref{lnregge1}, with the gap of order $L_\ell/\ell_P=4 \ln d[a]\gg 1$.

As a result, the dominant contribution of $\overline{\tr(\rho_A^n)}$ in Eq.\Ref{sum2} is given by
\be
\overline{\tr(\rho_A^n)}\sim%\prod_\fp\frac{1}{C_{n,\fp}}\lt(D^n\rt)^{N_\partial}
e^{-\frac{1}{8\pi \ell_P}\frac{V\lt(M_{n}\rt)}{L_{AdS}^2}+\text{boundary terms}}\label{dom}
\ee

The denominator $\overline{\tr(\rho_A)^n}$ in Eq.\Ref{Sn} can be computed in a very similar manner, since
\be
\tr(\rho_A)^n=\tr\lt[\lt(\rho_P^{\otimes n}\otimes_\fp|\cv_\fp\rangle\langle \cv_\fp|\rt)^{\otimes n}\rt].\label{tracerhon1}
\ee 
which different from Eq.\Ref{tracerhon} by removing $\cc^{(n)}_A$ in the trace. We still use the Haar random average Eq.\Ref{random}, and write $\overline{\tr(\rho_A)^n}$ as a sum over all permutations $\{g_{\fp}\}$ at all nodes. However because $\cc^{(n)}_A$ is absent, as $D\gg1$ the dominant configurations of $\{g_{\fp}\}$ satisfy the boundary condition that $g_\fp=I$ at the entire boundary. Thus suppose $\{g_{\fp}\}$ has different domains with different $g_\fp$, domain walls are detached from the boundary, and contain closed curves. 

Using the same argument as the above, we can write $\overline{\tr(\rho_A)^n}$ as a sum of path integral of Regge action on different $M_{\{g_\fp\}}$, similar to Eq.\Ref{sum2}. domain walls become the branch curves in $M_{\{g_\fp\}}$, which contain closed curves. However, since the intersection of two hyperbolic surfaces cannot give closed branch curves, $M_{\{g_\fp\}}$ with closed branch curves doesn't admit AdS$_3$ geometry. Thus the equation of motion doesn't have any solution, except $g_\fp=I$ identically without any domain wall. As a result $\overline{\tr(\rho_A)^n}$ is dominant at the configuration that $g_\fp=I$ everywhere
\be
\overline{\tr(\rho_A^n)}\sim%\prod_\fp\frac{1}{C_{n,\fp}}\lt(D^n\rt)^{N_\partial}
e^{-\frac{n}{8\pi \ell_P}\frac{V\lt({M}_{1}\rt)}{L_{AdS}^2}+\text{boundary terms}},
\ee
where the boundary terms are identical to the ones appearing in Eq.\Ref{dom}.

We find that the average R\'enyi entropy is given by 
\be
\overline{S_n(A)}\simeq \frac{1}{1-n}\lt[\ln Z(n)_\infty-n\ln Z(1)_\infty\rt]\label{duality}
\ee
where %up to a term $\ln\lt(\prod_\fp{\lt(D^n\rt)^{N_\partial}}/{C_{n,\fp}}\rt)$, 
\be
\ln Z(n)_\infty\equiv -\frac{1}{8\pi \ell_P}\frac{V\lt(M_{n}\rt)}{L_{AdS}^2}\label{Z(n)}
\ee
is the on-shell action of Einstein gravity on 3-manifold $M_n$. The relation Eq.\Ref{duality} has been an assumption in the existing derivation of RT formula from AdS/CFT \cite{Dong:2016fnf,dche,Lewkowycz:2013nqa}. But it is now derived from the state Eq.\Ref{Psi} using random tensor networks.

Inserting Eq.\Ref{lnregge1} in Eqs \Ref{duality} and \Ref{Z(n)}, we obtain the RT formula of R\'enyi entropy for CFT$_2$, which has the desired $n$ dependence.
\be
\overline{S_n(A)}\simeq \lt(1+\frac{1}{n}\rt)\frac{L_{AdS}}{8\ell_P}l_{\cs}\label{result}
\ee
where $L_{AdS}l_{\cs}$ corresponds to $\Ar_{min}$ the geodesic length in AdS$_3$ in Eq.\Ref{RT}. The usual RT formula is recovered as $n\to1$. The above result reproduces the Renyi entropy computed by Hung-Myers-Smolkin-Yale in \cite{Hung:2011nu} using the AdS/CFT assumptions. To see it is indeed the right R\'enyi entropy of the boundary CFT$_2$, recall the central charge of CFT relates to $L_{AdS}$ and $\ell_P$ by $c=\frac{3L_{AdS}}{2\ell_P}$, and $l_\cs$ relates to length $l_A$ of boundary interval $A$ by $l_\cs\simeq 2 \ln (l_A/\delta)$ in Poincar\'e patch, where $\delta$ is a UV cut-off. It gives
\be
\overline{S_n(A)}\simeq \lt(1+\frac{1}{n}\rt)\frac{c}{6} \ln \lt(\frac{l_A}{\delta}\rt)
\ee
which matches precisely the R\'enyi entropy Eq.\Ref{CFTS} of CFT$_2$ with correct $n$ dependence \cite{Calabrese:2004eu}.

\section{Conclusion and Outlook}

In this paper we explicitly construct a boundary state $|\Psi\rangle$ as a linear combination of random tensor networks to understand the holographic duality at the level of many-body quantum states. $|\Psi\rangle$ is shown to reproduce holographically the (1+1)d CFT R\'enyi entropies $S_n(A)$ with the correct $n$-dependence, which agrees with the known behavior of CFT ground states. 

The boundary state $|\Psi\rangle$ is given by the holographic mapping applied to the bulk discretized Wheeler-deWitt wave function $\Phi(\vec{a})$ as a path integral of (2+1)d gravity. The holographic CFT R\'enyi entropies are derived by the average of random tensors in the tensor networks. The random tensors effectively glue the path integrals and relate boundary R\'enyi entropies to partition functions of bulk gravity on branched cover spacetimes. The relation between boundary R\'enyi entropies and bulk gravity partition functions on branched cover spacetimes has been a key assumption in the existing holographic R\'enyi entropy computations \cite{Dong:2016fnf,dche,Lewkowycz:2013nqa}, while our work derives this relation with random tensor networks in the context of AdS$_3$/CFT$_2$. The derivation also gives the duality between the boundary R\'enyi entropy and bulk cosmic brane (in AdS$_3$/CFT$_2$), which has been proposed as a conjecture in \cite{Dong:2016fnf}. 

As another key feature, our derivation doesn't assume the bulk geometry corresponding to the entropy calculation. The geometries on branched cover manifolds are emergent from the boundary CFT state $|\Psi\rangle$ via the large bond dimension limit of tensor networks. $|\Psi\rangle$ is proposed as a new way to approximate the ground state of large-N CFT. 

The results in this paper suggest that $|\Psi\rangle$ is a promising realization of the holographic duality in the tensor network, which deserves to be further developed. It is interesting to generalize the derivation to higher dimensional CFTs, which is expected to derive the conjectured duality between the boundary R\'enyi entropy and bulk cosmic brane in general context. A current research undergoing is to understand the conformal symmetry of $|\Psi\rangle$. $|\Psi\rangle$ proposed as the CFT ground state is expected to be (approximately) invariant under boundary conformal transformations. The bulk Wheeler-deWitt wave function and its semiclassical behavior turns out to be a nice tool to understand the symmetry perspectives. It is also important to compare $|\Psi\rangle$ with the optimized MERA, given that both approaches captures the behavior of CFT ground state. The comparison may require to extend our analysis to the regime of finite bond dimensions (the finite-N CFT).  

The holographic mapping with random tensor networks should be applied to study the excited states in the boundary CFT. The states excited by certain boundary operators acting on $|\Psi\rangle$ should be the image of the holographic mapping from a bulk wave function, created by a bulk operator acting on the present Wheeler-deWitt wave function $\Phi(\vec{a})$. It suggests a holographic relation between boundary and bulk operators and is a promising starting point to understand the bulk reconstruction in the AdS/CFT \cite{Hamilton:2006az,Jafferis:2017tiu}. Since $\Phi(\vec{a})$ are the coefficients in the expansion of $|\Psi\rangle$ with random tensor networks, the relation between boundary and bulk operators should be an analog of the standard duality in quantum mechanics between operators acting on Dirac kets and operators acting on wave functions. %It is also interesting to formulate the present framework in the language of error correcting codes \cite{Almheiri:2014lwa,Harlow:2016vwg}, which understands systematically the reconstruction of bulk information from boundary excited states. 

This paper for the first time relates Regge calculus, as a popular tool in general relativity, to the tensor network, which is extensively studied in quantum information, condensed matter physics, and the AdS/CFT. The framework developed in this paper is a promising platform for communications between different fields. Regge calculus closely relates to loop quantum gravity, especially its covariant formulation \cite{semiclassical,CFsemiclassical,HZ,HHKR,Han:2017xwo,Kaminski:2017eew}. This work is an interesting starting point to explore the relation between loop quantum gravity, AdS/CFT correspondence, and string theory.

% Regge calculus

\begin{acknowledgements}
%The authors acknowledges the anonymous referees for their review of an earlier version of the manuscript and their insightful comments and suggestions. 

MH acknowledges Ling-Yan Hung and Yidun Wan at Fudan University in Shanghai, Yong-Shi Wu at University of Utah, Bei Zeng at University of Waterloo, and Jie Zhou at Perimeter Institute for Theoretical Physics for the hospitality during his visits, stimulating discussions, and collaborations. MH acknowledges the useful discussions with Xiao-liang Qi and Zhao Yang, and especially thanks Ling-Yan Hung for explaining the result in \cite{Hung:2011nu}. MH also acknowledges support from the US National Science Foundation through grant PHY-1602867, and the Start-up Grant at Florida Atlantic University, USA. 
\end{acknowledgements}

%%%%%%%%%%%%%%%%%%%%%%%%%%%%%%%%%%%%%%%%%%%%%%%%%%%%%%%%%%%%%%%%%%%%%%%%%%%%%%%%%%%%%%%%%%%%%%%%%%%%%%%%%%%%%%%%%%%%%%%%%%%%%%%

%%%%%%%%%%%%%%%%%%%%%%%%%%%%%%%%%%%%%%%%%%%%%%%%%%%%%%%%%%%%%%%%%%%%%%%%%%%%%%%%%%%%%%%%%%%%%%%%%%%%%%%%%%%%%%%%%%%%%%%%%%%%%%%

\appendix

\section{Bound on Fluctuation}\label{fluctuation}

In this appendix we examine the fluctuation of the R\'enyi entropy $S_n(A)$ from the average value $\overline{S_n(A)}$ in Eq.\Ref{result}, to qualify how well is the approximation. We show that in the regime $\ell_P\ll L_\ell$ the fluctuation is generically small. The method used in the following is similar to \cite{Qi1}.

We denotes by $Z(n)=\tr(\rho_A^n)$ and ${Z(n)}_\infty$ the average value of $Z(n)$ as $\ell_P\ll L_\ell$ (same as in Eq.\Ref{Z(n)}). We consider the following fluctuation of $Z(n)$:
\be
\overline{\lt(\frac{Z(n)}{{Z(n)}_\infty}-1\rt)^2}&=&\lt(\frac{\overline{Z(n)^2}}{{Z(n)^2_\infty}}-1\rt)-2\lt(\frac{\overline{Z(n)}}{{Z(n)_\infty}}-1\rt)\nonumber\\
&\leq&\lt(\frac{\overline{Z(n)^2}}{{Z(n)^2_\infty}}-1\rt)
\ee
$\overline{Z(n)}\geq {Z(n)_\infty}$ because in the approximation we made as $\ell_P\ll L_\ell$, the neglected terms in the sums are all non-negative.  

%Here $\ell_P\ll L_\ell$ means
%\be
%\ln d[a]\gg 1
%\ee
%$O(\ell_P)$ means $O(1/\ln d[a])$.

$\overline{Z(n)^2}$ is computed in a similar way as the above, using the random average formula Eq.\Ref{random}, changing $n$ by $2n$. It leads to that the dominant contribution of $\overline{Z(n)^2}$ is again given by a sum over permutations $\{g_\fp\}$ at all $\fp$, whose boundary condition is $g_\fp=I$ in $\bA$ and $g_{\fp}=(1\cdots n)(n+1\cdots2n)$ in $A$. Hence $\overline{Z(n)^2}$ is also written as a sum of path integrals on different $M_{\{g_\fp\}}$. The situation of a single domain wall separating $I$ in $R_\bA$ and $(1\cdots n)(n+1\cdots2n)$ in $R_A$ gives again the dominant contribution. The 3-manifold $M_{\{g_\fp\}}$ in this case is simply 2 copies of $M_n$. As a result,
\be
\frac{\overline{Z(n)^2}}{{Z(n)^2_\infty}}=\prod_\fp\frac{C_{2n,\fp}}{C_{n,\fp}^2}\lt[1+O\lt(\frac{\ell_P}{L_\ell}\rt)\rt],\quad \frac{C_{2n,\fp}}{C_{n,\fp}^2}\leq 1
\ee
which implies the following bound
\be
\overline{\lt(\frac{Z(n)}{{Z(n)}_\infty}-1\rt)^2}\leq  O\lt(\frac{\ell_P}{L_\ell}\rt)
\ee
Bounding the fluctuation of $Z(n)$ by $\eps/4$ has the following probability by Markov inequality,:
\be
\mathrm{Prob}\lt(\lt|\frac{Z(n)}{{Z(n)}_\infty}-1\rt|\geq\frac{\eps}{4}\rt)\leq \frac{\overline{\lt(\frac{Z(n)}{{Z(n)}_\infty}-1\rt)^2}}{\lt(\frac{\eps}{4}\rt)^2}\leq O\lt(\frac{\ell_P}{\eps^2L_\ell}\rt).
\ee
Similar conclusion can be drawn for $Z(1)^n$. Bounds on the fluctuations of $Z(n),Z(1)^n$ implies the bound on the fluctuation of $S_n(A)$. The probability of violating the following bound is of $ O\lt({\ell_P}/{\eps^2L_\ell}\rt)$
\be
\lt|S_n(A)-\overline{S_n(A)}\rt|&\leq& \frac{1}{n-1}\lt(\Bigg|\ln\frac{Z(n)}{{Z(n)}_\infty}\Bigg|+\lt|\ln\frac{Z(1)^n}{{Z(1)}^n_\infty}\rt|\rt)\nonumber\\
&\leq&\eps
\ee 
where we have used that $|\ln(1\pm \eps/4)|\leq\eps/2$ for small $\eps$. When $\ell_P/L_\ell\ll \eps^2$, the above bound of fluctuation is satisfied with a high probability $1-O\lt({\ell_P}/{\eps^2L_\ell}\rt)$.

\section{Compute $\overline{\tr(\rho_A^n)}$}\label{compute}

This appendix demonstrates a graphic computation of the random average $\overline{\tr(\rho_A^n)}$. It is sufficient to demonstrate the case of $n=2$, while the generalization to $n>2$ is obvious.

\begin{figure}[h]
\begin{center}
\includegraphics[width = 8cm]{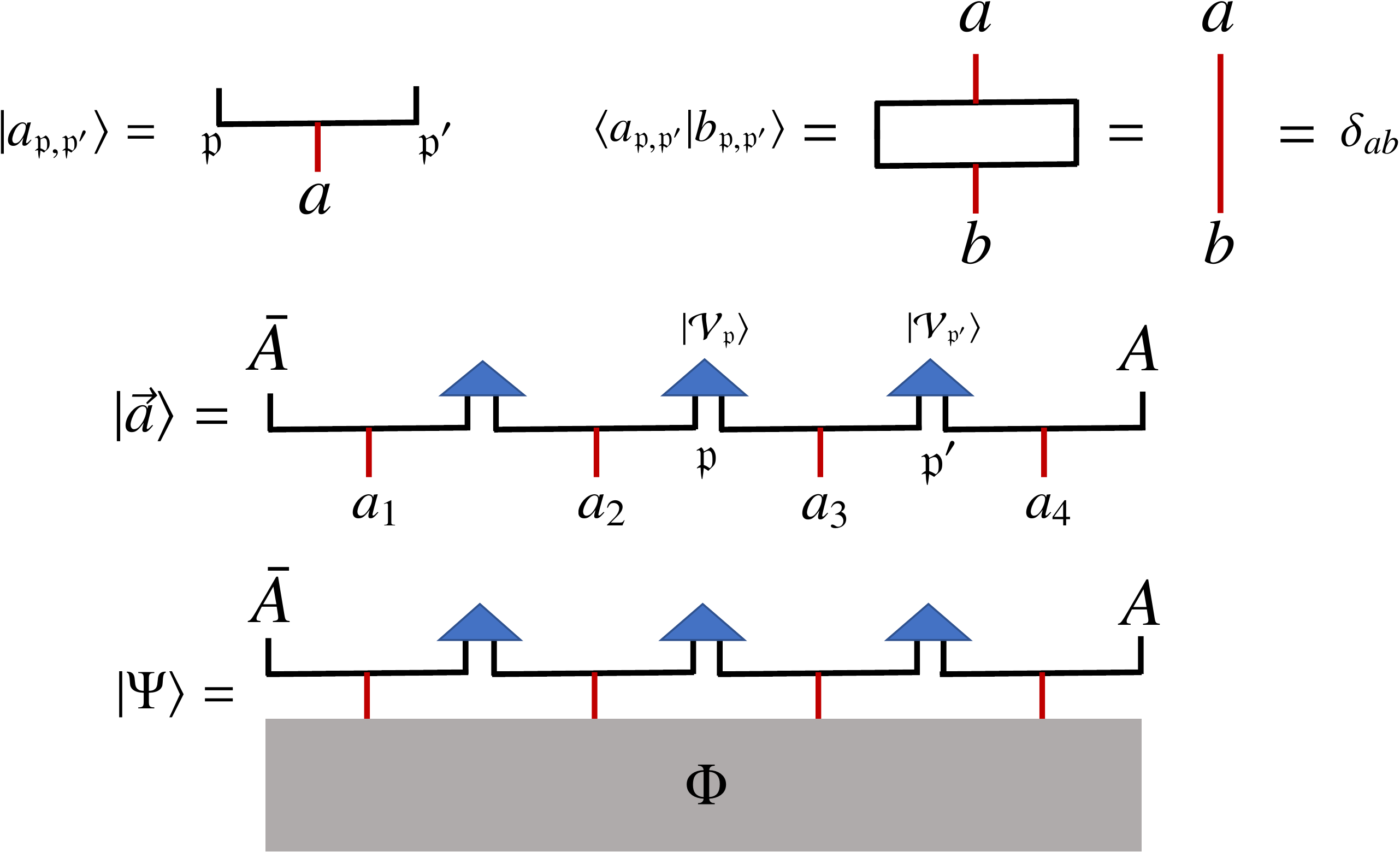}
\end{center}
\caption{Graphic presentation of the link state $|a_{\fp,\fp'}\rangle$, the inner product $\langle a_{\fp,\fp'}|b_{\fp,\fp'}\rangle$, the tensor network $|\vec{a}\rangle$, and $|\Psi\rangle$. Blue triangles in $|\vec{a}\rangle$ are random tensors $|\cv_{\fp}\rangle$ located at nodes $\fp$. Connecting the tensor network to $\Phi(\vec{a})$ (gray box) with red lines illustrate the sum over $\vec{a}$ in defining $|\Psi\rangle$. The gray box illustrates $\Phi(\vec{a})$ as the path integral over $M$.} 
\label{formula}
\end{figure}

FIG.\ref{formula} lists the basic ingredients of presenting graphically the computation of $\overline{\tr(\rho_A^2)}$. We have suppress 1 spatial dimension in the presentation where the tensor network $|\vec{a}\rangle$ is illustrated by a chain, and two ends of the chain illustrate regions $A$ and $\bA$.

\begin{figure}[h]
\begin{center}
\includegraphics[width = 0.36\textwidth]{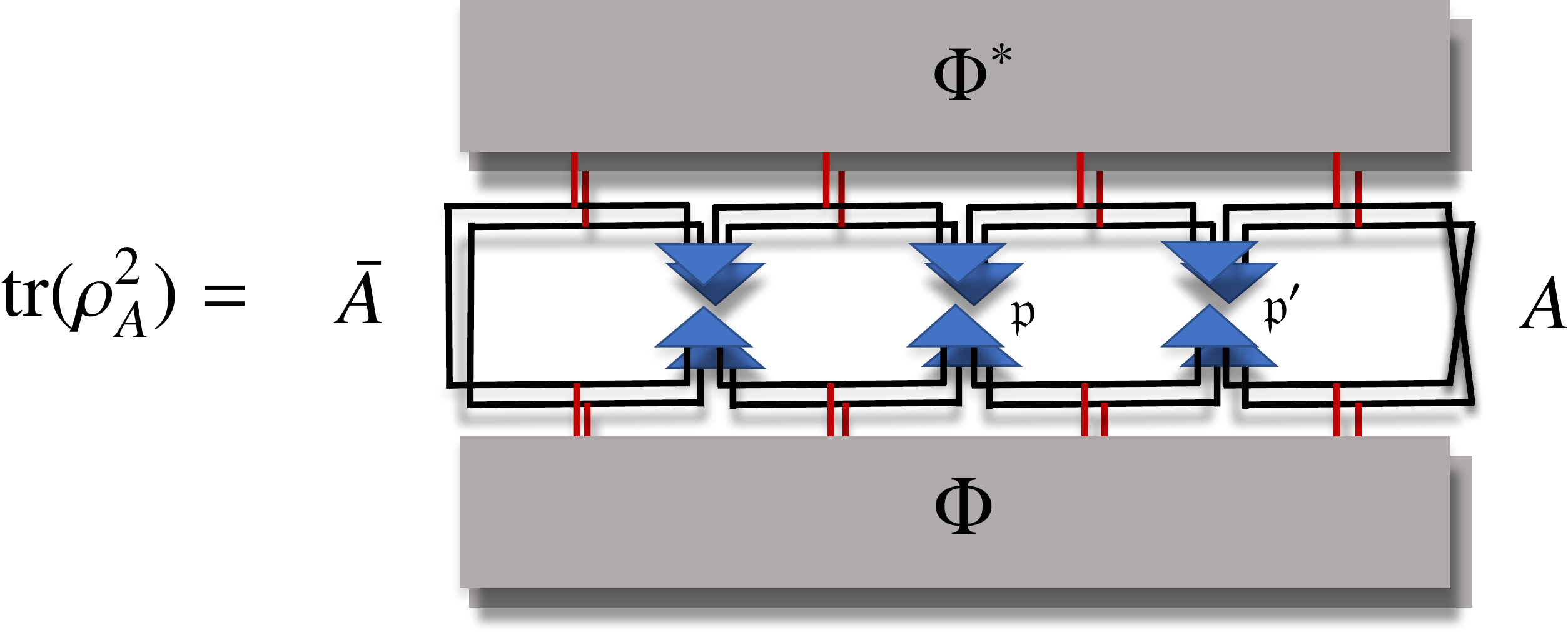}
\end{center}
\caption{Graphic illustration of $\tr(\rho_A^2)$.} 
\label{second}
\end{figure}

FIG.\ref{second} is a graphic illustration of $\tr(\rho_A^n)$ in Eqs.\Ref{tracehon1} and \Ref{tracerhon} at $n=2$. $\tr(\rho_A^n)$ contains $n$ copies of $\Phi^*(\vec{a})$ and $n$ copies of $\Phi(\vec{a})$.

\begin{figure}[h]
\begin{center}
\includegraphics[width = 8cm]{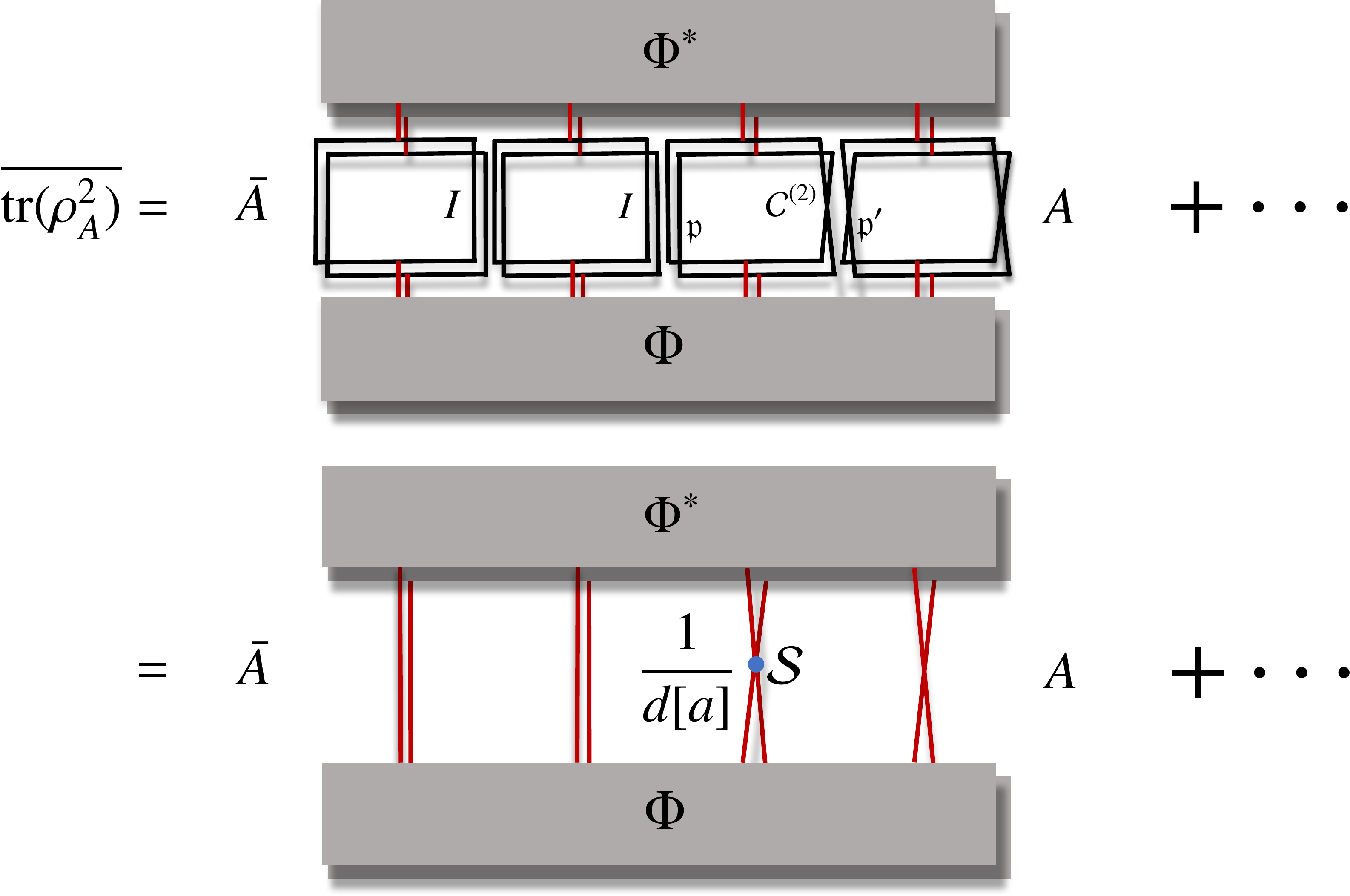}
\end{center}
\caption{The graphical computation of the average $\overline{\tr(\rho_A^2)}$.} 
\label{average}
\end{figure}

\begin{figure}[h]
\begin{center}
\includegraphics[width = 5cm]{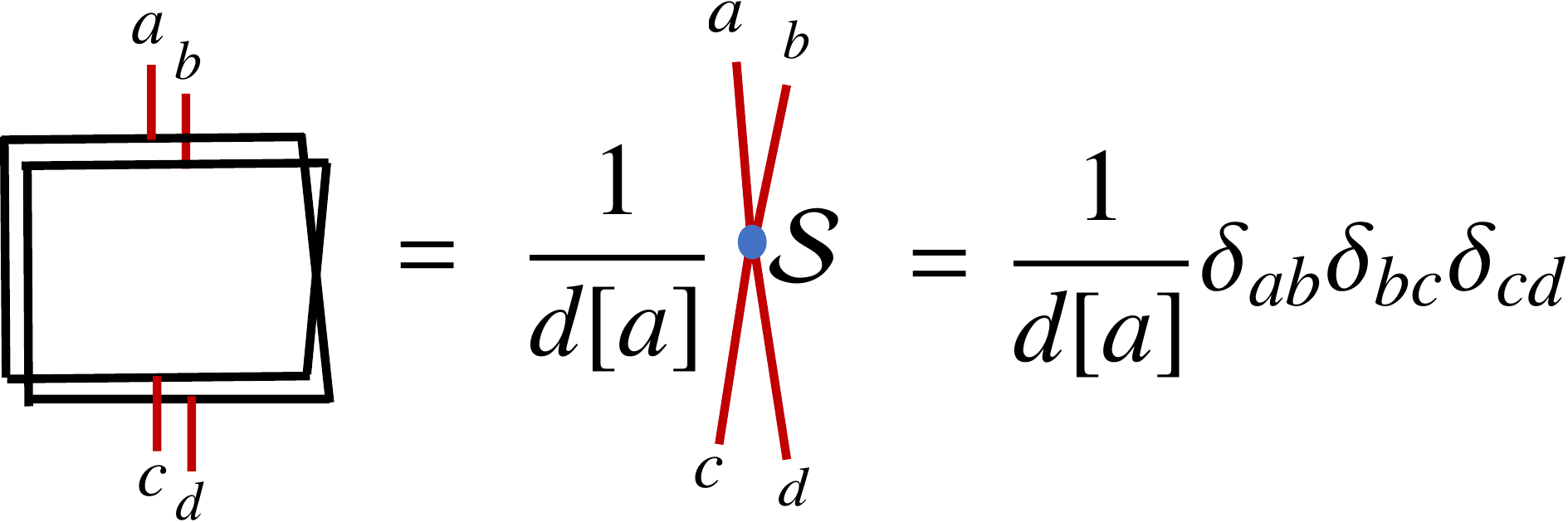}
\end{center}
\caption{The graphical computation involving inner products among 4 link states $|a_{\fp,\fp'}\rangle$, $|b_{\fp,\fp'}\rangle$, $|c_{\fp,\fp'}\rangle$, $|d_{\fp,\fp'}\rangle$ at a domain wall $\cs$. The involved link data $a,b,c,d$ are all identified at $\cs$, which makes $\cs$ a branch point (curve) of the branch cover manifold. The inner products give a prefactor $d[a]^{-1}$.  } 
\label{cross}
\end{figure}

The random average Eq.\Ref{random} is performed for $2n$ copies of random tensors locally at each node $\fp$. Applying Eq.\Ref{random} to $\overline{\tr(\rho_A^n)}$ gives a sum over choices of permutations $\{g_\fp\}$ at all $\fp$. The computation of $\overline{\tr(\rho_A^2)}$ is illustrated in FIG.\ref{average}, which shows a typical term corresponding to two domains with $g_\fp=I$ and $g_{\fp'}=\cc^{(2)}$. The second step in FIG.\ref{average} uses the relations in FIG.\ref{formula} and FIG.\ref{cross}. The red lines illustrate the identification and summation of the boundary data $\vec{a}$ in path integrals $\Phi(\vec{a}),\Phi(\vec{a})^*$, which glues the path integrals. The resulting path integral is on the branch cover manifold in FIG.\ref{replica}, where the branch points are in the domain wall $\cs$. Each link $(\fp,\fp')$ crossing $\cs$ associates a normalization factor $d[a]^{-1}$, generalized to $d[a]^{1-n_c}$ for an arbitrary pair $g_\fp,g_{\fp'}$, where $n_c$ is the length of the cycle $c$ in $g_\fp^{-1}g_{\fp'}$.

\section{Domain Wall Contributions}\label{DW}

\begin{figure}[h]
\begin{center}
\includegraphics[width=6cm]{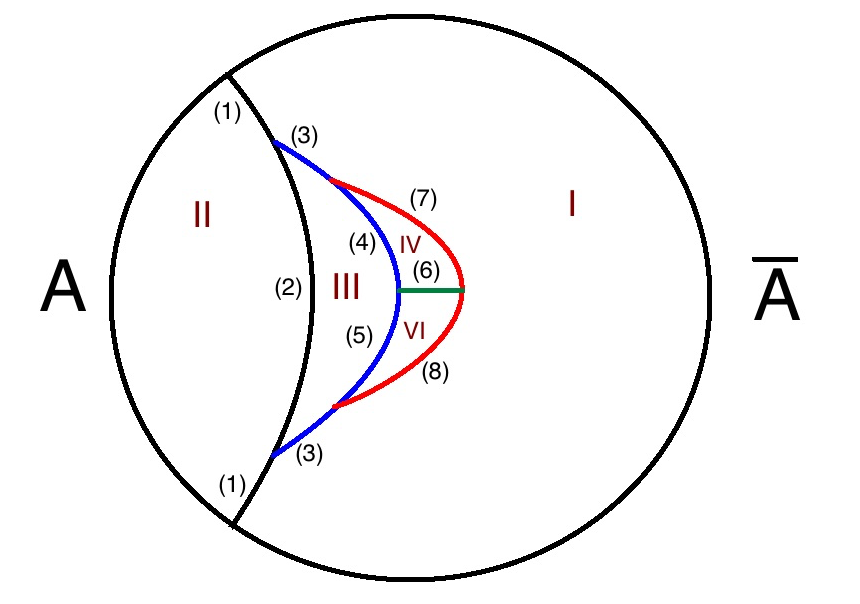}
\caption{shows the space $\Sig$ with boundary $\partial\Sig$ divided into regions $A$ and $\bA$. $\Sig$ contains the domain-walls $(1),(2),\cdots, (8)$, which divide the bulk of $\Sig$ into regions I, II, $\cdots$, VI. Each bulk region associates a permutation $g_{I, II, \cdots, VI}$, with $g_I=I$ and $g_{II}=(\cc^{(n)})^{-1}$.}
\label{walls}
\end{center}
\end{figure}

In this section, we prove that the configuration $\{g_\fp\}$ with a single domain wall indeed gives the leading contribution to $\sum_{\{g_\fp\}}$. Let's consider a more generic case shown in FIG.\ref{walls}(a), where more than one domain-walls are created in the bulk of $\Sig$. We are going to show that this configuration always contribute less than a single domain-wall.

Recall the computation of $V(M_{\{g_\fp\}})$ and the discussion below Eq.\Ref{26}. When we sum all domain walls and all cycles,  
\be
\frac{-V\lt(M_{\{g_\fp\}}\rt)}{8\pi \ell_P L_{AdS}^2}
= \frac{-n V\lt({M}_{1}\rt)}{8\pi \ell_P L_{AdS}^2}+\sum_{\cs_{g,g'}}\sum_{c\in C(g^{-1}g')}\frac{1-n_c^2}{n_c}\frac{L_{AdS}}{8\ell_P}l_{\cs_{g,g'}}.
\label{lnregge}
\ee
Thus each domain-wall carries the contribution proportional to
\be
l_{\cs_{g,g'}}\sum_{c\in C(g^{-1}g')}\frac{1-n_c^2}{n_c}\label{chiA}
\ee
in Eq.\Ref{lnregge} where $l_\cs$ is the geodesic length on the hyperbolic plane with unit curvature. Given trivalent intersection of domain-walls, which separate three domains with permutations $g_{1},g_2,g_3$ (FIG.\ref{flow}), we have the following ``triangle inequality'' (see Appendix \ref{proof} for a proof)
\be
\sum_{c\in C({g}_1^{-1}{g}_3)}\frac{1-n_c^2}{n_c}\geq \sum_{c\in C({g}_1^{-1}{g}_2)}\frac{1-n_c^2}{n_c}+\sum_{c\in C({g}_2^{-1}{g}_3)}\frac{1-n_c^2}{n_c}.\label{triineq}
\ee
It implies each trivalent intersection gives the following contribution
\be
&&l_{\cs_{13}}\sum_{c\in C({g}_1^{-1}{g}_3)}\frac{1-n_c^2}{n_c}+l_{\cs_{12}}\sum_{c\in C({g}_1^{-1}{g}_2)}\frac{1-n_c^2}{n_c}+l_{\cs_{23}}\sum_{c\in C({g}_2^{-1}{g}_3)}\frac{1-n_c^2}{n_c}\nonumber\\
%&\leq&l_{\cs_{13}}\sum_{c\in C({g}_1^{-1}{g}_3)}\frac{1-n_c^2}{n_c}+ \lt(\sum_{c\in C({g}_1^{-1}{g}_2)}\frac{1-n_c^2}{n_c}+\sum_{c\in C({g}_2^{-1}{g}_3)}\frac{1-n_c^2}{n_c}\rt)\mathrm{min}\lt(l_{\cs_{12}},l_{\cs_{23}}\rt)\nonumber\\
&\leq&\sum_{c\in C({g}_1^{-1}{g}_3)}\frac{1-n_c^2}{n_c}\lt[l_{\cs_{13}}+\mathrm{min}\lt(l_{\cs_{12}},l_{\cs_{23}}\rt)\rt]\nonumber
\ee
which is less than a single domain wall contribution.

For any intersection with e.g. 4 domain walls, one can always shift the end point of one domain wall away from the intersection and obtain a smaller $l_{\cs}$ (greater contribution to Eq.\Ref{lnregge}). It reduces the 4-valent intersection back to trivalent situation, which implies the contribution after the above shift is still smaller than the single domain wall configuration. The same argument applies to the intersection with larger number of domain walls.

Therefore we find that the contribution of the multi-domain-wall configuration is less or equal to the single domain-wall configuration
\be
l_{\cs_{g,g'}}\sum_{c\in C(g^{-1}g')}\frac{1-n_c^2}{n_c}\leq  \frac{1-n^2}{n}l_{\cs}.
\ee

\begin{figure}[h]
\begin{center}
\includegraphics[width=5cm]{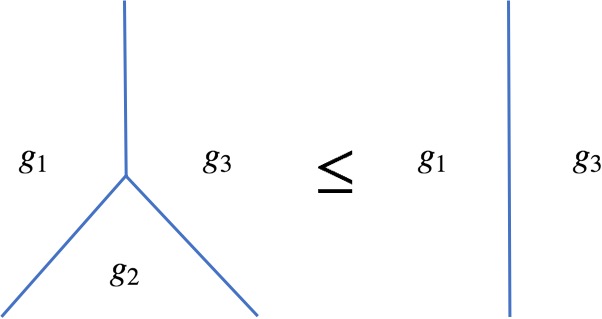}
\caption{A trivalent intersection of domain walls has less contribution than a single domain wall.}
\label{flow}
\end{center}
\end{figure}

\section{Proof of triangle inequality \Ref{triineq}}\label{proof}

For every permutation $g \in \text{Sym}_n$ (in the following discussion, we assume that $n \ge 2$), we can decompose $g$ as 
\[
	g = \prod_{i=1}^{k} c_i, 
	\]
where $c_i \in \text{Sym}_n$ are disjoint cycles such that $\sum_{i=1}^{k} n_{c_i} = n$. Denote $C(g) = \left\{ c_1, \ldots, c_k \right\}$ as the set of disjoint cycles whose product is $g$.

Let $d: \text{Sym}_n \rightarrow \mathbb{R}$ be a function which satisfies that there exists a function $f \in C^{2}[1,+\infty)$ such that (i) $f''(x) \le 0$ for $x \ge 1$; (ii) $\left(\frac{f(x)}{x}\right)' \ge 0$ for $x \ge 1$; (iii) For each permutation $g \in \text{Sym}_n$, we have $d(g) = \sum_{c \in C(g)} f(n_c)$.
We say $d$ is a norm on $\text{Sym}_n$, and $f$ is the {generator} of $d$. 

Let $f$ be a generator of a norm $d$ on $\text{Sym}_n$, then $f(1) = 0$. Indeed let $g = (1)(2 \ldots n)$, then $d(g) = f(1) + f(n-1) = f(1) + d(g)$, hence $f(1) = 0$.

We know $f'(x) \ge 0$, because leting $g(x) = f(x)/x$, we have $f'(x) = [x\cdot g(x)]' = x \cdot g'(x) + g(x) \ge g(x) = \int_{1}^{x} g'(x) \ge 0$.

Let $f$ be a generator of a norm on $\text{Sym}_n$. For every $x_1, \ldots x_k \ge 1$, we have 	\begin{eqnarray*}
		\sum_{i=1}^{k} f(x_i) = \sum_{i=1}^{k} x_i \frac{f(x_i)}{x_i} \le \sum_{i=1}^{k} x_i \frac{f(\sum_{i=1}^{k} x_i)}{\sum_{i=1}^{k} x_i} = f(\sum_{i=1}^{k} x_i).
	\end{eqnarray*}

Let $f$ be a generator of a norm on $\text{Sym}_n$. For every $x_1, \ldots x_k \ge 1$, we have $\sum_{i=1}^{k} f(x_i) \ge f(\sum_{i=1}^{k} x_i - k + 1)$. Indeed, when $k = 2$, we have
	\begin{eqnarray*}
		& f(x_1) + f(x_2) = \large\int_{1}^{x_1} f'(x) dx + \int_{1}^{x_2} f'(x) dx\\
		\ge & \int_{1}^{x_1} f'(x) dx + \int_{x_1}^{x_1+x_2-1} f'(x) dx = f(x_1 + x_2 - 1).
	\end{eqnarray*}
Suppose the argument holds when $k = k'$. When $k = k'+1$, we have
	\[  \sum_{i=1}^{k'+1} f(x_i) \ge f(\sum_{i=1}^{k'} x_i - k' + 1) + f(x_{k'+1}) \ge f(\sum_{i=1}^{k'+1} x_i - k').
		\]

Let $d: \text{Sym}_n \rightarrow \mathbb{R}$ be a norm on $\text{Sym}_n$ whose generator is $f$, $g \in \text{Sym}_n$ be a permutation, $c \in \text{Sym}_n$ be a cycle. Then we have $d(cg) \le d(c) + d(g)$. This statement can be proven by the following:

	Let $A \subseteq C(g)$ be the set of all cycles in $C(g)$ that are disjoint with $c$, Then $A \subseteq C(cg)$. Let $B_1 = C(g) \setminus A$, $B_2 = C(cg) \setminus A$. We obtain $ d(c) + d(g) - d(cg)=  f(n_c) + \sum\limits_{r \in B_1} f(n_r) - \sum\limits_{r \in B_2} f(n_r)$. Then $N = \sum_{r \in B_1} n_c = \sum_{r \in B_2} n_c$ leads to $\sum\limits_{r \in B_2} f(n_r) \le f(N)$. By the fact that $|B_1| \le n_c$, we have $\sum\limits_{r \in B_1} f(n_r) \ge f(N - |B_1| + 1) \ge f(N - n_c + 1)$. Therefore
	\[
		d(c) + d(g) - d(cg) \ge f(n_c) + f(N - n_c + 1) - f(N) \ge 0.
		\]

Let $d: \text{Sym}_n \rightarrow \mathbb{R}$ be a norm on $\text{Sym}_n$ whose generator is $f$. For $g_1, g_2 \in S_n$, we have $d(g_{1}g_{2}) \le d(g_1) + d(g_2)$. Indeed, let ${C}(g_1) = \left\{ c_1, \ldots, c_k \right\}$. We have
	\begin{eqnarray*}
		d(g_{1}g_{2}) = d((\prod_{i=1}^{k} c_i)g_2) \le d(c_1) + d((\prod_{i=2}^{k} c_i)g_2)\\
		\le \ldots \le \sum_{i=1}^{k} d(c_i) + d(g_2) =  d(g_1) + d(g_2).
	\end{eqnarray*}

$f(x) = \frac{x^2-1}{x}$ is a generator, since $\left(\frac{f(x)}{x}\right)' = \frac{2}{x^3} \ge 0$ when $x \ge 1$, and $ f''(x) = -\frac{2}{x^3} \le 0.$

As a result the function $d: \text{Sym}_n \rightarrow \mathbb{R}, d(g) = \sum_{c \in C(g)} \frac{1-n_{c}^2}{n_c}$ is a norm on $\text{Sym}_n$.

%\nocite{*}

%\bibliography{muxin}

\end{document}